\documentclass[%
 reprint,
superscriptaddress,
 amsmath,amssymb,
 aps,
prc,
]{revtex4-2}

\usepackage{graphicx}
\usepackage{dcolumn}
\usepackage{bm}
\usepackage{color}
\usepackage{hyperref}

\usepackage{braket}
\usepackage{multirow}
\usepackage{tabularx}

\begin{document}

\preprint{APS/123-QED}

\title{Isolated mixed-symmetry $2^+$ state of the radioactive neutron-rich nuclide $^{132}$Te}

\author{T. Stetz}
 \email{tstetz@ikp.tu-darmstadt.de}
 \affiliation{Technische Universität Darmstadt, Department of Physics, Institute for Nuclear Physics, 64289 Darmstadt, Germany}
 
\author{H. Mayr}
 \affiliation{Technische Universität Darmstadt, Department of Physics, Institute for Nuclear Physics, 64289 Darmstadt, Germany}

 \author{V. Werner}
 \email{vw@ikp.tu-darmstadt.de}
 \affiliation{Technische Universität Darmstadt, Department of Physics, Institute for Nuclear Physics, 64289 Darmstadt, Germany}

\author{N. Pietralla}
 \affiliation{Technische Universität Darmstadt, Department of Physics, Institute for Nuclear Physics, 64289 Darmstadt, Germany}

\author{Y. Tsunoda}
 \affiliation{Center for Nuclear Study, University of Tokyo, Hongo, Bunkyo-ku, Tokyo 113-0033, Japan}

\author{T. Otsuka}
 \affiliation{Center for Nuclear Study, University of Tokyo, Hongo, Bunkyo-ku, Tokyo 113-0033, Japan}

\author{G. Rainovski}
 \affiliation{Faculty of Physics, St. Kliment Ohridski University of Sofia, 1164 Sofia, Bulgaria}

\author{T. Beck}
 \affiliation{Technische Universität Darmstadt, Department of Physics, Institute for Nuclear Physics, 64289 Darmstadt, Germany}

\author{R. Borcea}
 \affiliation{Horia Hulubei National Institute of Physics and Nuclear Engineering, 077125 Bucharest-M\u{a}gurele, Romania}
 
\author{S. Calinescu}
 \affiliation{Horia Hulubei National Institute of Physics and Nuclear Engineering, 077125 Bucharest-M\u{a}gurele, Romania}

\author{C. Costache}
 \affiliation{Horia Hulubei National Institute of Physics and Nuclear Engineering, 077125 Bucharest-M\u{a}gurele, Romania}

\author{I. E. Dinescu}
 \affiliation{Horia Hulubei National Institute of Physics and Nuclear Engineering, 077125 Bucharest-M\u{a}gurele, Romania}

\author{K. Gladnishki}
 \affiliation{Faculty of Physics, St. Kliment Ohridski University of Sofia, 1164 Sofia, Bulgaria}

\author{K. E. Ide}
 \affiliation{Technische Universität Darmstadt, Department of Physics, Institute for Nuclear Physics, 64289 Darmstadt, Germany}

\author{A. N. Ionescu}
 \affiliation{Horia Hulubei National Institute of Physics and Nuclear Engineering, 077125 Bucharest-M\u{a}gurele, Romania}

\author{D. Kocheva}
 \affiliation{Faculty of Physics, St. Kliment Ohridski University of Sofia, 1164 Sofia, Bulgaria}

\author{P. Koseoglou}
 \affiliation{Technische Universität Darmstadt, Department of Physics, Institute for Nuclear Physics, 64289 Darmstadt, Germany}

\author{R. Lica}
 \affiliation{Horia Hulubei National Institute of Physics and Nuclear Engineering, 077125 Bucharest-M\u{a}gurele, Romania}

\author{N. M\u{a}rginean}
 \affiliation{Horia Hulubei National Institute of Physics and Nuclear Engineering, 077125 Bucharest-M\u{a}gurele, Romania}

\author{C. Mihai}
 \affiliation{Horia Hulubei National Institute of Physics and Nuclear Engineering, 077125 Bucharest-M\u{a}gurele, Romania}

\author{R. E. Mihai}
 \affiliation{Horia Hulubei National Institute of Physics and Nuclear Engineering, 077125 Bucharest-M\u{a}gurele, Romania}
 \affiliation{Institute of Experimental and Applied Physics, Czech Technical University in Prague, Husova 5, Prague, Czech Republic}
 
\author{C. M. Nickel}
 \affiliation{Technische Universität Darmstadt, Department of Physics, Institute for Nuclear Physics, 64289 Darmstadt, Germany}

\author{C. R. Nita}
 \affiliation{Horia Hulubei National Institute of Physics and Nuclear Engineering, 077125 Bucharest-M\u{a}gurele, Romania}

\author{L. Stan}
 \affiliation{Horia Hulubei National Institute of Physics and Nuclear Engineering, 077125 Bucharest-M\u{a}gurele, Romania}

\author{S. Toma}
 \affiliation{Horia Hulubei National Institute of Physics and Nuclear Engineering, 077125 Bucharest-M\u{a}gurele, Romania}

\author{R. Zidarova}
 \affiliation{Technische Universität Darmstadt, Department of Physics, Institute for Nuclear Physics, 64289 Darmstadt, Germany}

\date{\today}

\begin{abstract}
The $M1$ transition strengths between excited $2^+$ states of the neutron-rich, radioactive nuclide $^{132}$Te have been studied through direct lifetime measurements using the Doppler-shift attenuation method in a two-neutron transfer reaction on a $^{130}$Te target. An unambiguous identification of the lowest-lying mixed-symmetry $2^+$ state has been achieved on the basis of the large $B(M1;2^+_2$\,$\rightarrow$\,$2^+_1$)\,=\,0.18(2)\,$\mu_\mathrm{N}^2$ transition strength. Results are compared to the shell model, and the analysis of both, data and calculations, unambiguously identifies the second-excited $2^+$ state of $^{132}$Te as the one-quadrupole phonon mixed-symmetry state of this isotope. A lowering of the energy and $B(M1;2^+_\mathrm{ms}\rightarrow 2^+_1)$ strength within the $N$\,=\,80 isotones toward the $Z$\,=\,50 shell closure is observed, which goes alongside with the lowering of the $E2$ collectivity approaching the magic proton shell.
\end{abstract}

\maketitle

\section{\label{sec:level1}Introduction}

Atomic nuclei are prime examples of a strongly correlated, two-component many-body quantum system. This fundamental character results in the prominent features of nuclei that are manifested in their shell structure, their capability of supporting collective phenomena, and their isospin degree of freedom. The delicate interplay and balance between these aspects in the formation of the nuclear excitation schemes and of nuclear decay reactions make the quantitative understanding of nuclear structure still a challenge. 

For a quantitative modeling of nuclear structure in wide regions of the nuclear chart, it is advisable to benchmark the accuracy of the description of its building blocks. The strongest channel of low-energy nuclear collectivity is given by the quadrupole degree of freedom as evidenced by the occurrence of a first excited $J^\pi\,=\,2^+$ state for almost all even-even nuclei in the nuclear chart. 
An intuitive, and yet quantitative, formulation of these building blocks of valence-shell quadrupole collectivity has been provided by the interacting-boson model with proton bosons and neutron bosons (IBM-2) \cite{lecturenotes,arima}. Due to the strong proton-neutron quadrupole-quadrupole interaction, it supports the formation of isoscalar quadrupole-collective structures at low excitation energies and of isovector valence-shell excitations at higher energies. The isoscalar structures are fully symmetric with respect to the pairwise exchange of proton-boson and neutron-boson labels while the other contain at least one pair of proton and neutron bosons which is antisymmetric under pairwise exchange of their isospin labels. These states are addressed as having {\em mixed symmetry} \cite{OTSUKA19781} and form an entire class of nuclear states with little mixing with isoscalar structures \cite{Iachello}. In vibrational even-even nuclei, the lowest-lying mixed-symmetry state is experimentally known \cite{mspietralla} to be the one-quadrupole phonon $2^+_{\mathrm{1,ms}}$ state. In a simplified, two configuration mixing scenario the lowest-energy one-quadrupole phonon states may be expressed as

\begin{align}
\label{eq:fs}
    Q_s \ket{0^+_1} = \ket{2^+_\mathrm{1}} \ \ \ &= \ \ \, \alpha \ket{2^+_\pi} + \beta \ket{2^+_\nu} \ , \\
\label{eq:ms}
    Q_m \ket{0^+_1} = \ket{2^+_\mathrm{1,ms}} &= -\beta \ket{2^+_\pi} + \alpha \ket{2^+_\nu} \ .
\end{align}

Both states are formed by the isoscalar $Q_s = Q_\pi + Q_\nu$ and mixed-symmetry $Q_m = Q_\pi/N_\pi - Q_\nu/N_\nu$ quadrupole operators acting on the $0^+$ ground state. Hence, they are fundamental one-phonon quadrupole excitations. $N_{\pi(\nu)}$ is the number of proton (neutron) bosons, equal to 1, each, for $^{132}$Te. 

While the term phonon is typically connected to vibrational excitation schemes, in the following we apply it also to the underlying microscopic configurations, which are the building blocks of what may become more collective phonon excitations in systems with larger numbers of valence particles. This approach allows to extend the (collective) mixed-symmetry concept to nuclei near double-magic shell closures, and to stay consistent with the symmetric and mixed-symmetric phonon labels outlined in literature \cite{mspietralla}. We note, that this analogy and the use of the term phonons does not imply the presence of developed vibrational structures in near-magic nuclei.

In proximity to a doubly-magic shell closure, the underlying proton and neutron $2^+$ configurations are often identified with the first-excited $2^+$ states of the nearest isotopes with only two valence protons or neutrons, respectively \cite{heydesau,werner,holt,covello}. Due to the isovector nature of the $2^+_\mathrm{1,ms}$ state, it decays with a strong $M1$ transition to the $2^+_1$ state, with a matrix element in the order of 1\,$\mu_\mathrm{N}$. This decay property of the $2^+_\mathrm{1,ms}$ state is the key signature for its identification. In addition, the ground-state transition of the $2^+_\mathrm{1,ms}$ state is known to be a weakly-collective $E2$ transition, which usually amounts to approximately 1\,W.u. \cite{mspietralla}.

The formation of isoscalar and mixed-symmetry quadrupole excitations in the vicinity of double shell closures is a particularly intriguing object of nuclear structure studies because they represent the building blocks of proton and neutron contributions to the formation of quadrupole collectivity in the corresponding regions of the nuclear chart \cite{stuch}. The isovector character of the $2^+_{\mathrm{1,ms}}$ state is then manifested in its preferential decay by magnetic dipole radiation. While in some cases the $2^+_\mathrm{1,ms}$ state has been assigned on the basis of a small multipole mixing ratio for its $E2/M1$-mixed decay to the $2^+_1$ state \cite{ham84}, an unambiguous identification of the $2^+_\mathrm{1,ms}$ state can only be achieved by the determination of the comparatively large $\bra{2^+_1}M1\ket{2^+_\mathrm{1,ms}}$ matrix element. Reliable and established methods to achieve this absolute measurement are Coulomb excitation or direct lifetime measurements. Both require a significant counting statistics that had been a challenge for the case of neutron-rich radioactive nuclei that are particularly interesting for benchmarking nuclear structure modeling for astrophysical applications. Due to the development of experimental methods over recent years, the study of mixed-symmetry states in neutron-rich radioactive nuclides has attracted much interest \cite{danchev,moschner,kern}.

In general, the above-mentioned quadrupole phonons are defined by action of the isoscalar and isovector $E2$ operators on a given state. Near shell closures often configurations of the type $(j^2)^{(L)}$, for two particles in an orbital $j$ coupling to total angular momentum $L$, are found to dominate the wave functions of low-lying states. These configurations form the building blocks of precursors of vibrational phonons in the above sense, which has extensively been discussed in particular near the $N$\,=\,50 shell closure (see Refs. \cite{werner,zr,wex,holt,casp}).

With the introduction of shell stabilization \cite{rain}, the $N$\,=\,80 isotonic chain, with two neutron holes with respect to the $N$\,=\,82 shell closure, has become a center of attention in the field of mixed-symmetry states. Information about the $2^+_\mathrm{1,ms}$ state in these isotones accumulated rapidly, reaching down to $^{132}$Te with only two protons above the $Z$\,=\,50 proton shell closure. This isotope is of particular interest because it is close and isobaric to the doubly-magic $^{132}$Sn. With only two valence protons and two valence-neutron holes, the mixing of the respective proton and neutron configurations into the wave functions is of utmost importance, as it constitutes the foundation for more complex collective states when turning to larger valence spaces. In addition, it makes a microscopic description within the nuclear shell model in large valence spaces possible.

The isotope $^{132}$Te has previously been investigated in a Coulomb-excitation experiment at the Holifield Radioactive Ion Beam Facility (HRIBF) at Oak Ridge National Laboratory (ORNL), with a $^{132}$Te beam impinging on a $^{12}$C target. The absolute $B(M1;2^+_2$\,$\rightarrow$\,$2^+_1$) strength has been determined through the state's population yields \cite{danchev}. Measurements of the magnetic moment of the $2^+_1$ state of $^{132}$Te at HRIBF/ORNL have shown a balance of proton and neutron contributions to the wave function of the $2^+_1$ state \cite{magmom1,magmom2,magmom3}. The same is expected for its mixed-symmetric counter part, the $2^+_\mathrm{1,ms}$ state. 
On the basis of the measured $M1$ strength, the $2^+_2$ state was assigned the most probable candidate for the $2^+_\mathrm{1,ms}$ state and using the literature branching ratio of $I(2^+_2$\,$\rightarrow$\,$2^+_1)/I(2^+_2$\,$\rightarrow$\,$0^+_1)$\,=\,100(52) from $\beta$-delayed $\gamma$-ray spectroscopy \cite{hughes}, a value of $B(M1;2^+_2$\,$\rightarrow$\,$2^+_1)\,=\,5.4(35)$\,$\mu_\mathrm{N}^2$ had been determined. However, this $B(M1)$ value stands out greatly among all measured $B(M1;2^+_\mathrm{1,ms}$\,$\rightarrow$\,$2^+_1)$ strengths throughout the nuclear chart. Its significant uncertainty is due to the large relative uncertainty of the $2^+_2$ state's decay branching ratio, and due to the low statistics in the $2^+_2$\,$\rightarrow$\,$2^+_1$ transition in Ref. \cite{danchev} which, at the time, represented the limit of experimental accessibility at an early radioactive-beam facility. Within the same study, however, a lower limit for the transition strength has been presented, which does not include the literature values of the branching ratio, but rather a lower limit deduced from the detection limit of the Coulomb-excitation data itself. This lower limit of $B(M1;2^+_2$\,$\rightarrow$\,$2^+_1) > 0.23$\,$\mu_\mathrm{N}^2$ was found to be in agreement with shell model calculations resulting in $B(M1;2^+_2$\,$\rightarrow$\,$2^+_1)\,=\,0.2$\,$\mu_\mathrm{N}^2$. At this point it can be concluded that the $2^+_2$ is the most probable candidate for the $2^+_\mathrm{1,ms}$ state of $^{132}$Te, however, an unambiguous assignment can only be made from a precise determination of the $2^+_2$\,$\rightarrow$\,$2^+_1$ transition strength. For this purpose a direct lifetime measurement of the $2^+_2$ state of the radioactive nuclide $^{132}$Te has been performed within the present work.



\section{Experiment}
The lower limit of the $2^+_2$\,$\rightarrow$\,$2^+_1$ transition strength from the Coulomb-excitation data implies that the lifetime of the $2^+_2$ state must be in the order of a few hundred femtoseconds. The Doppler-shift attenuation method is suitable to determine lifetimes in this range. This method is based on the energy shift of an emitted $\gamma$ ray, which depends on the velocity and the angle of observation with respect to the momentum vector of the recoiling nucleus. For this purpose, a reaction mechanism which produces fast-moving $^{132}$Te ions in their excited state is needed. Furthermore, in order to minimize feeding effects, the reaction mechanism is desired to favor direct population of the $2^+_2$ state. A two-neutron transfer reaction satisfies these requirements and was carried out with a runtime of approximately 300 hours at the 9\,MV FN Pelletron tandem accelerator at IFIN-HH in Bucharest-M\u{a}gurele, Romania. An $^{18}$O beam at 59\,MeV kinetic energy impinged on a $^{130}$Te target, populating excited states of $^{132}$Te via the $^{130}$Te($^{18}$O,$^{16}$O)$^{132}$Te two-neutron transfer reaction and at the same time inducing Coulomb excitation of the $^{130}$Te target nuclei. The beam energy was chosen to be slightly below the Coulomb barrier of the projectile-target system to ensure maximum yields in the two-neutron transfer channel in comparison to fusion-evaporation events. The beam had an average current of approximately 9\,pnA. The target with a thickness of 2\,mg/cm$^2$ and an enrichment to 99.4\,\% in $^{130}$Te was deposited on an 11\,mg/cm$^2$ thick $^{181}$Ta backing.

The Doppler-shift attenuation method makes use of the detection of deexcitation radiation at large forward and backward polar angles with respect to the direction of movement of the ions. The ROSPHERE detector array is divided into five rings at polar angles of 37$^{\circ}$, 70$^{\circ}$, 90$^{\circ}$, 110$^{\circ}$, and 143$^{\circ}$ with respect to the beam axis \cite{rosphere}. 
ROSPHERE was equipped with 25 high-purity germanium (HPGe) detectors, individually surrounded by bismuth germanium oxide (BGO) detectors for Compton suppression. The BGO detectors were not shielded by heavy-metal collimators in order to facilitate their use as multiplicity detectors. The trigger condition of the data acquisition was set to the detection of a $\gamma$ ray by one or more HPGe detectors in coincidence with an event recorded by one or more solar cells of the SORCERER particle detector \cite{sorcerer}. The particle condition is necessary because even at the chosen beam energy, the number of fusion-evaporation events exceeded the desired two neutron-transfer reaction by more than one order of magnitude. The ejectile of the two-neutron transfer reaction is predominantly scattered in the backward direction, hence, the SORCERER array equipped with 18 silicon photo diodes of about 40\,mm$^2$ in size, each, in a 2-ring configuration was placed in that position, covering polar angles of 109°\,$< \theta <$\,163° with respect to the beam axis \cite{sorcerer}.

\section{Analysis}
In the transfer reaction, excited states of $^{132}$Te were populated and, due to the reaction kinematics, the Te ions were recoiling in the forward direction within an angular spread of less  than 5$^{\circ}$. Furthermore, excited states of the target materials $^{130}$Te and $^{181}$Ta were populated by Coulomb excitation, as well as excited states of $^{134}$Xe via $\alpha$ transfer from $^{18}$O. The kinematics of all three reactions were indistinguishable by the utilized suite of particle detectors and, hence, all are present in the recorded particle-$\gamma$ data. The recoiling ions, with an initial average velocity of about 0.02\,$c$, were continuously slowed down within the target and backing materials until coming to rest. This stopping process is simulated using the program \texttt{StopSim} \cite{apcad}. 
The stopping powers of the ions within the target materials, which are needed for the simulation of the velocity distribution, 
have been calculated with the program \texttt{SRIM} \cite{srim,srim2}. Depending on the velocity of the isotopes at the moment of deexcitation, the energy of the $\gamma$ radiation will be shifted to higher (lower) energies in forward (backward) direction with respect to the beam. This results in a line shape of the observed peak in the spectrum detected at the corresponding positions. These line shapes contain larger Doppler-shifted fractions with increasing average velocity, i.e., shorter lifetimes, of the ions in their excited states, and with deviation of the angle of observation from 90$^{\circ}$.
The line shape can, hence, be fitted with the lifetime of the excited state as a free parameter. This method is limited by the time it takes the isotopes to come to rest, as no shift will occur afterwards. The line-shape analysis has been performed using the computer code \textsc{APCAD} \cite{apcad}. The data were sorted into spectra for the individual rings, i.e., groups of detectors sharing a given polar angle $\theta$. The program is able to fit multiple peaks and line shapes in all rings, simultaneously. 
Within this work, 
all presented values have been obtained with the \textsc{MIGRAD} minimizer and the statistical uncertainties have been provided by the \textsc{MINOS} algorithm \cite{minuit}. Systematic uncertainties result from the calculated electronic and nuclear stopping powers which have been varied in the simulation process by 5$\,\%$ and 10$\,\%$, respectively. The energy-dependent resolution of the HPGe detectors has been considered by fitting a detector-response function of the form $f(E)$\,=\,$a+b \sqrt{E}$ to the calibration data.

\begin{figure}[t]
\centering
\includegraphics[width=1.0\columnwidth]{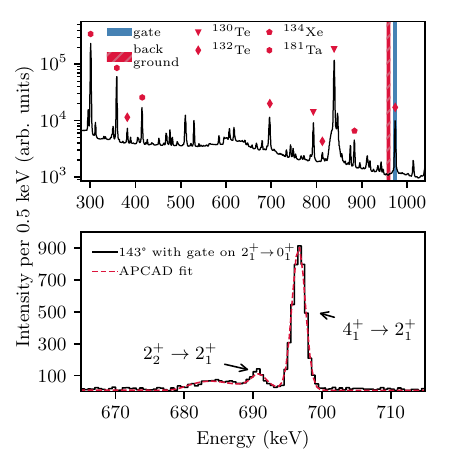}
\caption{\label{fig:143}  A particle-gated, background-subtracted spectrum, summed over all HPGe detectors is shown (top panel). Major transitions between excited states of the four reaction channels are indicated. The range of the coincidence condition for the line-shape analysis of $2^+ \rightarrow 2^+_1$ transitions is marked by the vertical bars. The energy spectrum of the 143$^{\circ}$ ring is shown (bottom panel) with a coincidence condition set on the ground state transition of the $2^+_1$ state in all detectors as shown in the top panel. The transition of the $4^+_1$ state to the $2^+_1$ state is present at 697\,keV. However, the two transitions are clearly separated and the Doppler shift of the $2^+_2$\,$\rightarrow$\,$2^+_1$ transition to lower energies in backwards direction is visible as a line shape on the left of the peak.}
\end{figure}

In the present experiment, the decays from the first three excited $2^+$ states of $^{132}$Te were observed, of which the $2^+_2$ and $2^+_3$ states showed line shapes due to their short lifetimes. The longer-lived $2^+_1$ state, which was to a high degree fed by higher-lying states, was used to define a particle-$\gamma$-$\gamma$ gating condition, in order to derive clean spectra for the $2^+_{2,3}$\,$\rightarrow$\,$2^+_1$ transitions. No transitions feeding these $2^+$ states of interest were observed. The resulting spectrum for the largest backward angle of 143$^{\circ}$ is shown in Fig. \ref{fig:143}. In addition to the $2^+_2$\,$\rightarrow$\,$2^+_1$ transition, the dominant peak in the depicted energy region is the transition of the $4^+_1$ state to the $2^+_1$ state. The two transitions are clearly separated, and no Doppler-shifted components were observed for the $4^+_1$\,$\rightarrow$\,$2^+_1$ decay. The fit resulting from the line-shape analysis of the $2^+_2$\,$\rightarrow$\,$2^+_1$ transition is included as a dashed line in Fig. \ref{fig:143}. At forward angles the $4^+_1$\,$\rightarrow$\,$2^+_1$ transition overlaps with the line shape of the desired transition of the $2^+_2$ state and has simultaneously been fitted as a stopped contaminant. 

To account for potential unobserved feeding into the states of interest, a multiplicity filter was implemented. A condition was set on the detection of exactly two $\gamma$ rays in two different HPGe detectors. Taking into account the total efficiency of the HPGe+BGO array of about 50\,\%, events with multiplicities\,$>$\,2, in particular $\gamma$ cascades feeding into the $2^+$ states of interest, are accordingly suppressed. The line-shape analysis was performed with and without the implementation of the multiplicity filter. The results showed no significant changes in the determined lifetime. Hence, the $2^+_2$ state was almost exclusively populated directly through the two-neutron transfer reaction. Nevertheless, a systematic uncertainty in the lifetime of the $2^+_2$ state of $\Delta_\mathrm{syst}\tau(2^+_2)$\,=\,0.05\,$\mathrm{ps}$ was estimated and added in quadrature to the dominating statistical uncertainty.  

The resulting lifetime of the $2^+_2$ state is given in Table \ref{tab:tau} along with the lifetime of the $2^+_3$ state which was determined accordingly. In the analysis process of the $2^+_3$\,$\rightarrow$\,$2^+_1$ transition at 813\,keV, a distinct contaminant was present at about 816\,keV, which originates from $\gamma$-ray transitions occurring in both $^{130}$Te and $^{132}$Te, and was taken into account in the fitting procedure as described above.

The ground-state transition of the $2^+_2$ state was not observed in this experiment. An upper limit for the branching ratio of the $2^+_2$\,$\rightarrow$\,$0^+_1$ transition was determined from the detection limit within a $3\,\sigma$ interval to $I_{2^+_2 \rightarrow 0^+_1}/I_{2^+_2 \rightarrow 2^+_1}$\,$<$\,2.4\,$\%$ in agreement with previous findings \cite{hughes}.

\begin{table}[b]
\caption{\label{tab:tau}%
Lifetimes of excited $2^+$ states obtained within this work are listed.
}
\begin{ruledtabular}
\begin{tabular}{ccc}
$J^\pi_i$ & Energy $E$ & Lifetime $\tau$ \\
\colrule
$2^+_2$ & 1665\,keV & 0.92(7)\,ps \\
$2^+_3$ & 1788\,keV & 7.6(27)\,ps
\end{tabular}
\end{ruledtabular}
\end{table}

The multipole mixing ratio of the $2^+_2$\,$\rightarrow$\,$2^+_1$ transition is yet unkown. However, the Coulomb excitation yields \cite{danchev} and the known branching ratio already indicate a predominant $M1$ character of the $2^+_2$\,$\rightarrow$\,$2^+_1$ transition. Considering the vibrational limit of $B(E2;2^+_2$\,$\rightarrow$\,$2^+_1$)\,$\leq$\,$2\times B(E2;2^+_1$\,$\rightarrow$\,$0^+_1$) yields a limit for the multipole mixing ratio of $\delta$\,$<$\,0.38 and, hence, a predominant fraction of $>87$ \% $M1$ multipolarity for this transition in any reasonable nuclear structure scheme.

From the obtained lifetimes, the limit on the multipole mixing ratio, and the literature decay branching ratio, transition strengths can be determined, which are given in Table \ref{tab:b}.

\begin{table}[b]
\caption{\label{tab:b}%
Transition strengths in units of $\mu_{\mathrm{N}}^2$ for $M1$ and W.u. for $E2$ transitions for the three lowest-lying $2^+$ states from the present work are listed and compared to literature data from Coulomb excitation (CoulEx) \cite{radford,danchev} and shell-model calculations, using effective charges and $g$ factors from Ref. \cite{BrownSN100PN} as discussed in the text. The ambiguity of the multipole mixing ratio of the $2^+_2$\,$\rightarrow$\,$2^+_1$ transition leads to the addition of a systematic uncertainty of 0.01$\,\mu_\mathrm{N}^2$ to the $M1$ transition strength. Branching ratios of the $2^+_3$ state were taken from \cite{nucldata}. Due to the unknown multipole mixing ratio for the $2^+_3$\,$\rightarrow$\,$2^+_1$ transition, upper limits are given for the respective $M1$ and $E2$ transition strengths.
}
\begin{ruledtabular}
\begin{tabular}{lrcccc}
$J_i$ & $J_f$ & $\pi\lambda$ & This work & CoulEx & SM \\
\colrule
$2^+_1$ & $ 0^+_1$ & $E2$ & & 10(1) & 8.7 \\
$2^+_2$ & $ 2^+_1$ & $M1$ & 0.18(2) & $>$0.23\footnote{\label{CoulOnly}from CoulEx only \cite{danchev}}\,,\,5.4(3.5)\footnote{\label{CoulDec}from CoulEx including literature branching ratio \cite{danchev,hughes}} & 0.27 \\
                           &  & $E2$ & $<$19.2 & $<$20 & 0.03 \\
        & $ 0^+_1$ & $E2$ & $<$0.04\footnote{\label{limit}using upper limit of the branching ratio from this work}, 0.017(9)\footnote{\label{withbranch}from this work including literature branching ratio \cite{hughes}} & 0.5(1)\textsuperscript{\ref{CoulDec}} & 0.02 \\
$2^+_3$ & $ 2^+_1$ & $M1$ & $<$0.013 & & 0.001 \\
                           &  & $E2$ & $<$7.2 & & 4.3 \\
        & $ 0^+_1$ & $E2$ & 0.061$^{+39}_{-19}$ & & 0.40 \\
\end{tabular}
\end{ruledtabular}
\end{table}


\section{Discussion}

\begin{figure}[t]
\centering
\includegraphics[scale=1]{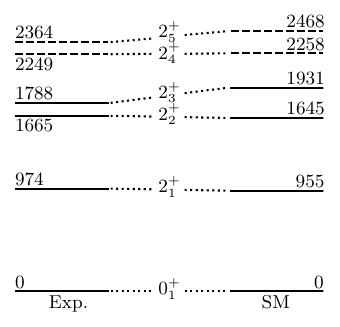}
\caption{\label{fig:level} Levelscheme of low-lying $2^+$ states of $^{132}$Te in comparison with the level energies obtained by shell-model calculations.}
\end{figure}

Given the large uncertainties on the literature values from Refs. \cite{radford,danchev}, the lifetimes for the $2^+_{2,3}$ states obtained in the present work give the first quantitative determination of the $M1$ decay strength of the first mixed-symmetry $2^+$ state of the radioactive, neutron-rich nuclide $^{132}$Te and first information on its potential fragmentation. With the directly-measured lifetime of the $2^+_2$ state, the largely uncertain, but absolutely small, $\gamma$-decay branching ratio of this state to the ground state plays nearly no role in the determination of its transition strength to the $2^+_1$ state. The size of the $M1$ matrix element amounts to $|\langle 2^+_1 \parallel M1 \parallel 2^+_2 \rangle | = 0.95(5) \mu_N$ in agreement with the value expected for a $2^+_{1, \mathrm{ms}}$ state. Also the systematic uncertainty due to the unknown multipole mixing ratio of the $2^+_2$\,$\rightarrow$\,$2^+_1$ decay, considering physics constraints, is small. In the following, the resulting $B(M1; 2^+_{2,3}$\,$\rightarrow$\,$2^+_1)$ strengths are compared to predictions from shell-model calculations. 

\begin{figure}[t]
\centering
\includegraphics[width=1.0\columnwidth]{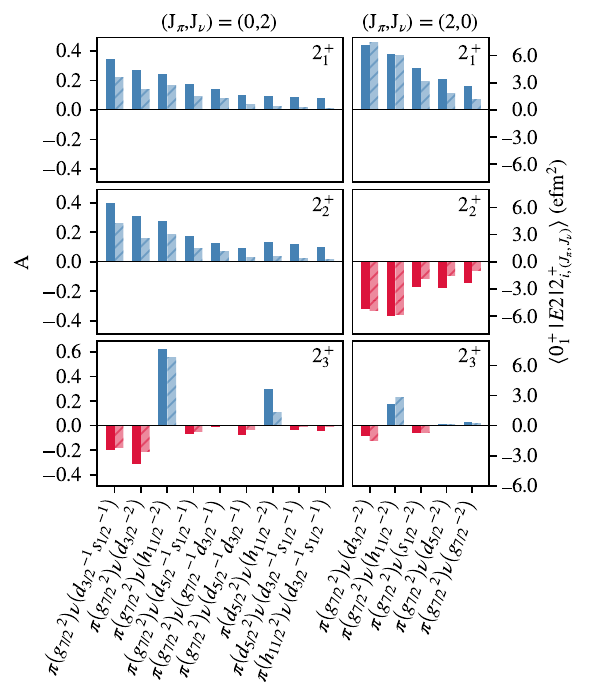}
\caption{\label{fig:ampl} Calculated amplitudes (left, solid bars) and their contributions to the $E2$ matrix elements to the ground state (right, hatched bars), respectively, are shown for the leading components of the $2^+_1$ (top), $2^+_2$ (middle), and $2^+_3$ (bottom) states. All components to the shell-model wave functions that contribute at least 1\,\% squared amplitude are shown. Wave function amplitudes with positive signs are marked blue and those with negative signs are marked red. The same color scheme applies to their contributions to the $E2$ matrix elements to the ground state.}
\end{figure}


These calculations were performed in a valence space composed of the $0g_{7/2}$, $1d_{5/2}$, $1d_{3/2}$, $2s_{1/2}$, and $0h_{11/2}$ orbitals for both, protons and neutrons, using the SN100PN interaction \cite{BrownSN100PN} and the code KSHELL \cite{kshell1,kshell2}. Proton and neutron effective charges were set to $e_\pi$\,=\,1.7\,$e$ and $e_\nu$\,=\,0.8\,$e$, which have more widely been used in this region (see, e.g., Refs \cite{gray2020,hicks2022}). The magnetic moment operator operator
\begin{equation}
    \mu_{\rm eff} = g_{l,{\rm eff}} {\bf l} + g_{s,{\rm eff}} {\bf s} + g_{p,{\rm eff}} [Y_2,{\bf s}] 
\end{equation}
and the corresponding effective $g_{\rho,{\rm eff}}$ factors have been adopted from Ref. \cite{BrownSN100PN}. 


The focus of this work is on the description of the lowest-lying $2^+$ states of interest. As shown in Figure \ref{fig:level}, the calculation reproduces well the energies of low-lying excited $2^+$ states of $^{132}$Te with the largest deviation below 10\,\% for the $2^+_3$ state. Furthermore, $M1$ and $E2$ transition strengths have been calculated and are included in Table \ref{tab:b}, along with the experimental values. 
The calculated $B(M1; 2^+_2$\,$\rightarrow$\,$2^+_1)$ transition strength agrees with the mixed-symmetry expectations, but exceeds the data by about 50 \%. The known $g$ factor of $g(2^+_1)\,=\,(+)0.46(5)$ \cite{magmom2,danchev} of the first excited state is well reproduced with $g(2^+_1)_\mathrm{SM}\,=\,0.48$. We note, however, that the choice of free orbital $g_l$ factors and standard spin $g_s$ factors with a quenching of 0.7 would improve the match of the calculated $B(M1; 2^+_2$\,$\rightarrow$\,$2^+_1)$ transition strength of 0.19\,$\mu_\mathrm{N}^2$ with data, while still resulting in a reasonably good $g(2^+_1)$ value of 0.40. 

The calculated $g$ factors of the first two excited $2^+$ states with $g(2^+_1)_\mathrm{SM}\,=\,0.48$ and 
$g(2^+_2)_\mathrm{SM}\,=\,0.36$ are rather similar and reflect the balance of proton and neutron contributions to their respective wave functions.

For the ground-state transition of the $2^+_2$ state, the calculations suggest a small $E2$ transition strength of 0.02\,W.u. The experimental value, calculated from the present lifetime measurement and the literature branching ratio, of only $0.02(1)$\,W.u. confirms this small $E2$ strength and is in good agreement with the shell model calculations. From the present data alone, i.e., without resorting to the highly uncertain literature value for its decay branching ratio into the ground state, we obtain an upper limit of 0.04\,W.u.

\begin{figure}[t]
\centering
\includegraphics[width=1.0\columnwidth]{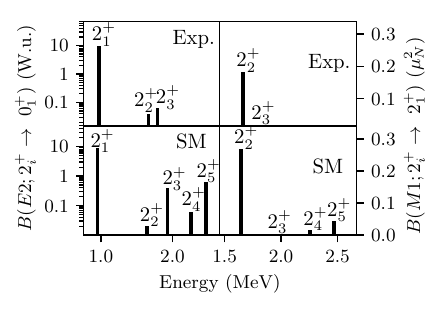}
\caption{\label{fig:e2m1} $B(E2;2^+_i$\,$\rightarrow$\,$0^+_1$) transition strengths are shown in l.h.s. panels (logarithmic scale), and $B(M1;2^+_{i>1}$\,$\rightarrow$\,$2^+_1$) transition strengths in r.h.s. panels (linear scale) for the three lowest-lying $2^+$ states. The upper panels show data obtained from Coulomb excitation \cite{danchev} and the present work. Analogously, the transition strengths obtained from shell-model calculations are shown for the five lowest-lying $2^+$ states in the lower panels.}
\end{figure}

The structure of the first three $2^+$ states is investigated in more detail within the shell-model approach by analysis of the amplitudes of leading wave function components in all three states. Figure \ref{fig:ampl} shows the leading neutron and proton configurations, indicated by the configurations of $J$\,=\,2 couplings. The $2^+_1$ and $2^+_2$ states are dominated by configurations in which either the proton pair or the neutron pair contributes to the spin of the respective state. These configurations are denoted as $(J_\pi,J_\nu)$, with values of $(0,2)$ or $(2,0)$. For these leading configurations, i.e., configurations which contribute to the wave function with $>$\,1\,$\%$ in either state, the absolute values in both, the first and second excited $2^+$ states are nearly identical. However, all leading $(0,2)$ components of the $2^+_{1,2}$ states' wave functions have the {\it same} sign, whereas all $(2,0)$ configurations have {\it opposite} signs. Therefore, the overlaps of $(0,2)$ and $(2,0)$ configurations in the $2^+_{1,2}$ states,
\begin{align*}
    \braket{2^+_1;(0,2)|2^+_2;(0,2)} & = +0.994 \approx +1 ,\\
    \braket{2^+_1;(2,0)|2^+_2;(2,0)} & = -0.976 \approx -1 ,
\end{align*}
confirm the isoscalar and isovector characters of their structures, respectively, which is expected for the symmetric $2^+_1$ state and the $2^+_{\mathrm{1,ms}}$ state from Eqs. (\ref{eq:fs}) and (\ref{eq:ms}) in a simple two-configuration mixing scheme. This clearly identifies the $2^+_{1,2}$ states as the one-phonon fully-symmetric and mixed-symmetry $2^+$ states of interest, respectively. This identification is corroborated by the corresponding coherent contributions of the proton and neutron parts to the $E2$ matrix elements of the $2^+_{1,2}$  states to the ground state. While both are in phase for the $2^+_1 \rightarrow 0^+_1$ transition and constructively add up to an appreciable $E2$ strength of calculated 8.7\,W.u., they are out of phase for the $2^+_2 \rightarrow 0^+_1$ transition and interfere destructively to a negligibly small $E2$ strength of a fraction of a W.u. For both $2^+_{1,2}$ states the proton and neutron parts of their wave functions can be considered as collective, because they are fragmented over many configurations coherently adding to the $E2$ matrix elements, and they are almost identical between both states, except for a relative phase between proton and neutron contributions. This supports the notion of the $2^+_{1,\mathrm{ms}}$ state as a collective isovector valence shell excitation \cite{lecturenotes,arima,OTSUKA19781,Iachello}.

\begin{figure}[t]
\centering
\includegraphics[width=1.0\columnwidth]{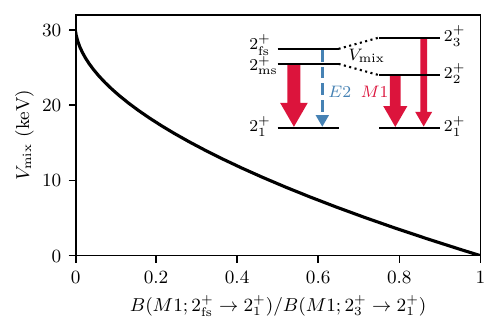}
\caption{\label{fig:vmix}The mixing matrix element V$_\mathrm{mix}$ between the $2^+_2$ state and $2^+_3$ state is shown as a function of the assumed $M1$ transition strength from a fully symmetric $2^+$ state to the $2^+_1$ state. By decreasing $B(M1;2^+_\mathrm{fs}$\,$\rightarrow$\,$2^+_1$), an upper limit for the mixing matrix element can be obtained. The $B(M1;2^+_\mathrm{fs}$\,$\rightarrow$\,$2^+_1$) value is limited by the $M1$ transition strength of the small fragment, which in this case corresponds to $B(M1;2^+_3$\,$\rightarrow$\,$2^+_1$)\,$<$\,0.013\,$\mu_\mathrm{N}^2$. The mixing scenario is schematically shown in the top right corner.}
\end{figure}

The measurement of the $2^+_3$ lifetime also allows to investigate the potential mixing between fully-symmetric and mixed-symmetric wave function components and the resulting possible fragmentation of the $2^+_{\mathrm{1,ms}}$ state over multiple $2^+$ states. Such fragmentation has been observed in the $N$\,=\,80 isotones at higher $Z$, in particular, in $^{138}$Ce \cite{rain}, where it had been attributed to the closure of the proton $0g_{7/2}$ orbital and the corresponding lack of shell stabilization of the $2^+_{\mathrm{1,ms}}$ wave function \cite{rain}. In the neighboring isotopes, $^{136}$Ba \cite{barium} and $^{140}$Nd \cite{kern,williams,Gladnishki}, the mixed-symmetry configuration was found to be predominantly concentrated in a single state, similar to isotopes near the $N$\,=\,50 neutron shell \cite{zr,mo,ru}. The most likely fragmentation should occur over states in proximity of the $2^+_{\mathrm{1,ms}}$ main component, most importantly the $2^+_3$ state. The results from Table \ref{tab:b} show a nearly negligible $M1$ strength connecting the $2^+_3$ and $2^+_1$ states, hence, the mixed-symmetry strength appears to be concentrated in the second excited $2^+$ state of $^{132}$Te. This experimental result is corroborated by the shell-model calculations, which also give a very small value for the $M1$ component of the $2^+_3$\,$\rightarrow$\,$2^+_1$ transition, despite of its proximity in energy within 123 keV, while at the same time its $E2$ component agrees with the limits set by the data. Shell-model wave-function amplitudes of the $2^+_3$ state are included in Figure \ref{fig:ampl} and show no similarities to the patterns of the $2^+_{1,2}$ states. Further $2^+$ states were included in the calculations, however, they are considerably higher in energy than the $2^+_3$ state and none of these states has a significant $M1$ transition strength to the $2^+_1$ state as shown in Figure \ref{fig:e2m1}.

Similarly to Ref. \cite{kern} we applied a mixing calculation between unperturbed fully-symmetric and mixed-symmteric $2^+$ configurations to the $2^+_2$ and $2^+_3$ states. The mixing calculation requires information about the typical $M1$ strength between fully-symmetric $2^+$ states, which is expected to be close to zero but is experimentally unknown. Nevertheless, already the $B(M1;2^+_3$\,$\rightarrow$\,$2^+_1$) strength is lower than the $M1$ transition strengths of assumed fully-symmetric $2^+_\mathrm{fs}$ states to the $2^+_1$ state in neighboring isotopes \cite{ahn,danchev,kern}. Therefore, we performed a series of two-state mixing calculations in which the $B(M1;2^+_\mathrm{fs}$\,$\rightarrow$\,$2^+_1$) strength was varied with an upper limit of $B(M1;2^+_\mathrm{fs}$\,$\rightarrow$\,$2^+_1)$\,$\leq$\,$B(M1;2^+_3$\,$\rightarrow$\,$2^+_1$). The resulting mixing matrix element V$_\mathrm{mix}$ is given in Figure \ref{fig:vmix} as a function of the assumed $M1$ strengths between fully-symmetric states. This analysis gives an upper limit for the mixing matrix element of $V_\mathrm{mix}$\,$\leq$\,31\,keV, however, we note that this limit applies to the idealized case of vanishing $M1$ transitions between fully-symmetric states, and an even lower value of $V_{\mathrm{mix}}$ is likely for more realistic assumptions. This finding corroborates the notion of mixed-symmetry states as an entire class of states \cite{OTSUKA19781,Iachello}, the wave functions of which can robustly persist with little mixing in an environment of other low-energy excitations. 

\begin{figure}[t]
\centering
\includegraphics[width=1.0\columnwidth]{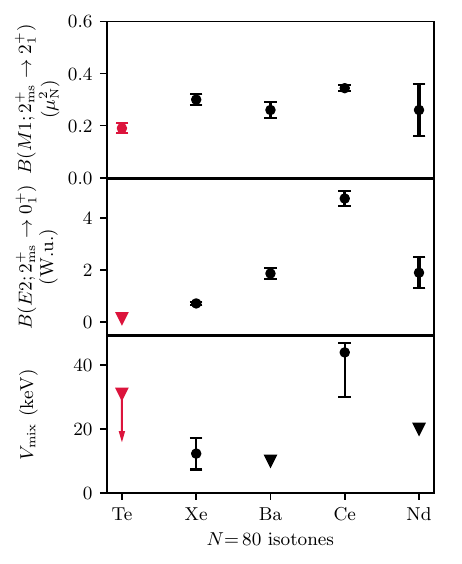}
\caption{\label{fig:m1e2v} $B(M1;2^+_\mathrm{ms}$\,$\rightarrow$\,$2^+_1$) (top) and $B(E2;2^+_\mathrm{ms}$\,$\rightarrow$\,$0^+_1$) (middle) transition strengths, respectively their upper limits (triangle), along the $N$\,=\,80 isotonic chain are shown. In the case of $^{138}$Ce, summed transition strengths of the $2^+_{2-6}$ states are shown. The V$_\mathrm{mix}$ (bottom) reflects the fragmentation of the $M1$ strength observed in $^{138}$Ce \cite{rain}. Data are taken from \cite{ahn,barium,rain,kern}.}
\end{figure}

The present result very well fits into the trend previously observed for the $N$\,=\,80 isotones \cite{kern}, shown in Figure \ref{fig:m1e2v}. The mixing matrix element is relatively small between about 10\,keV and 20\,keV \cite{ahn,barium}, which is a sign for the purity of the $2^+_{\mathrm{1,ms}}$ states. In contrast to this, it maximizes at Ce \cite{rain} where the closure of the $\pi(g_{7/2})$ proton sub-shell at proton number $Z$\,=\,58 leads to a stronger fragmentation of one-quadrupole phonon configurations and, thus, to a larger mixing matrix element between them. This trend is accompanied by the systematic behaviors of $M1$ and $E2$ transition strengths. The $E2$ strengths from the mixed-symmetry $2^+$ states drop toward the $Z$\,=\,50 shell closure with a minimum at $^{132}$Te. The $B(M1;2^+_\mathrm{ms} \rightarrow 2^+_1)$ strengths are rather constant in the open proton shell, summing multiple $2^+_{\mathrm{1,ms}}$ fragments in the case of $^{138}$Ce \cite{rain}, and slightly drop in the system with only two valence protons and two valence neutrons, $^{132}$Te. The one-phonon mixed-symmetry state being concentrated in the second-excited $2^+$ states, and the trend of diminishing $M1$ strengths toward a two-proton and two-neutron system, have previously been observed only for $^{92}$Zr \cite{werner,holt}, with two protons above $Z$\,=\,38 and two neutrons above $N$\,=\,50. However, in that case the special valence space with a low-lying proton $p_{1/2}$ orbital led to an asymmetric distribution of proton and neutron wave function components over the lowest two $2^+$ states and to the phenomenon of Configurational Isospin Polarization \cite{werner,holt} of the corresponding wave functions. For $^{132}$Te, no such effect is present, as also evidenced by known $2^+_1$ magnetic moments \cite{magmom1,magmom2,magmom3}.

\section{Conclusion}

The lifetimes of the second- and third-excited $2^+$ states of the neutron-rich, radioactive nuclide $^{132}$Te was measured using the Doppler-shift attenuation method following a two-neutron transfer reaction on a $^{130}$Te target. The resulting lifetime of the $2^+_2$ state of 0.92(7)\,ps resolves ambiguities from literature, and the deduced $B(M1;2^+_2 \rightarrow 2^+_1)$ strength clearly identifies this state as the main fragment of the one-phonon mixed-symmetry $2^+$ state of $^{132}$Te. Shell model calculations agree with the new data and a wave function analysis shows the expected signatures for the clear mixed-symmetry character of the $2^+$ state. A drop of its $E2$ excitation strength and its $M1$ strength connecting to the fully-symmetric $2^+_1$ state as compared to its heavier isotones is evident in this isotope with only two valence protons and neutron holes, each. Hence, the observed states constitute the building blocks of proton-neutron collective fully-symmetric and mixed-symmetry quadrupole-excitations in this minimal valence space. From the simultaneous measurement of the $2^+_3$ state little mixing between wave functions with mixed symmetry and those with full proton-neutron symmetry is found. Given a larger fragmentation of this isovector valence-shell excitation in the $N$\,=\,84 isotones $^{142}$Ce and $^{144}$Nd \cite{vanhoy1995,hicks1998} it is important to obtain similar data on the $N$\,=\,84 counterpart of $^{132}$Te with two valence protons and two valence neutrons outside of $^{132}$Sn. Lifetimes of excited $2^+$ states of $^{136}$Te will allow to pin down the underlying basic proton and neutron configurations and their coupling to the respective symmetric and mixed-symmetric eigenstates. Corresponding information will allow to assess the building blocks of quadrupole-collectivity for the region of neutron-rich radioactive nuclides located ”north-east” of $^{132}$Sn in the nuclear chart which is important for the modelling and understanding of the $r$-process path of cosmic nucleosynthesis \cite{RevModPhys.93.015002}.

\begin{acknowledgments}
The experiment was carried out in 2021 during the peak of the COVID-19 pandemic. For that reason, travel to the experiment was not possible and sincere thanks are due to the colleagues at IFIN-HH who professionally performed all hands-on work during the experiment. Also the staff of the IFIN-HH tandem accelerator is acknowledged for providing optimal beam conditions and for target preparation. This work was supported by the German BMBF under Grant No. 05P21RDCI2, by the Deutsche Forschungsgemeinschaft (DFG, German Research Foundation) – Project-ID 499256822 – GRK 2891 'Nuclear Photonics', by Project ELI-RO/DFG/2023\_001 ARNPhot funded by the Institute of Atomic Physics, Romania, and by the Romanian Nucleu project No. PN 23 21 01 02. K.G., D.K. and G.R. acknowledge support by the European Union-NextGenerationEU, through the National Recovery and Resilience Plan of the Republic of Bulgaria, project No. BG-RRP-2.004-0008-C01, and by
the Bulgarian Ministry of Education and Science, within the National Roadmap for Research Infrastructures (object CERN) and T.O. support by the German Alexander von Humboldt-Foundation.
The data that support the findings of this article are not publicly
available. The data are available from the authors upon reasonable
request.

\end{acknowledgments}



\begin{thebibliography}{44}%
\makeatletter
\providecommand \@ifxundefined [1]{%
 \@ifx{#1\undefined}
}%
\providecommand \@ifnum [1]{%
 \ifnum #1\expandafter \@firstoftwo
 \else \expandafter \@secondoftwo
 \fi
}%
\providecommand \@ifx [1]{%
 \ifx #1\expandafter \@firstoftwo
 \else \expandafter \@secondoftwo
 \fi
}%
\providecommand \natexlab [1]{#1}%
\providecommand \enquote  [1]{``#1''}%
\providecommand \bibnamefont  [1]{#1}%
\providecommand \bibfnamefont [1]{#1}%
\providecommand \citenamefont [1]{#1}%
\providecommand \href@noop [0]{\@secondoftwo}%
\providecommand \href [0]{\begingroup \@sanitize@url \@href}%
\providecommand \@href[1]{\@@startlink{#1}\@@href}%
\providecommand \@@href[1]{\endgroup#1\@@endlink}%
\providecommand \@sanitize@url [0]{\catcode `\\12\catcode `\$12\catcode
  `\&12\catcode `\#12\catcode `\^12\catcode `\_12\catcode `\%12\relax}%
\providecommand \@@startlink[1]{}%
\providecommand \@@endlink[0]{}%
\providecommand \url  [0]{\begingroup\@sanitize@url \@url }%
\providecommand \@url [1]{\endgroup\@href {#1}{\urlprefix }}%
\providecommand \urlprefix  [0]{URL }%
\providecommand \Eprint [0]{\href }%
\providecommand \doibase [0]{https://doi.org/}%
\providecommand \selectlanguage [0]{\@gobble}%
\providecommand \bibinfo  [0]{\@secondoftwo}%
\providecommand \bibfield  [0]{\@secondoftwo}%
\providecommand \translation [1]{[#1]}%
\providecommand \BibitemOpen [0]{}%
\providecommand \bibitemStop [0]{}%
\providecommand \bibitemNoStop [0]{.\EOS\space}%
\providecommand \EOS [0]{\spacefactor3000\relax}%
\providecommand \BibitemShut  [1]{\csname bibitem#1\endcsname}%
\let\auto@bib@innerbib\@empty
\bibitem [{\citenamefont {Iachello}(1976)}]{lecturenotes}%
  \BibitemOpen
  \bibfield  {author} {\bibinfo {author} {\bibfnamefont {F.}~\bibnamefont
  {Iachello}},\ }\bibfield  {title} {\bibinfo {title} {Lecture notes on
  theoretical physics}} (\bibinfo {year} {1976}),\ \bibinfo {note}
  {$\mathrm{R}$ijksuniversiteit Groningen}\BibitemShut {NoStop}%
\bibitem [{\citenamefont {Arima}\ \emph {et~al.}(1977)\citenamefont {Arima},
  \citenamefont {Ohtsuka}, \citenamefont {Iachello},\ and\ \citenamefont
  {Talmi}}]{arima}%
  \BibitemOpen
  \bibfield  {author} {\bibinfo {author} {\bibfnamefont {A.}~\bibnamefont
  {Arima}}, \bibinfo {author} {\bibfnamefont {T.}~\bibnamefont {Ohtsuka}},
  \bibinfo {author} {\bibfnamefont {F.}~\bibnamefont {Iachello}},\ and\
  \bibinfo {author} {\bibfnamefont {I.}~\bibnamefont {Talmi}},\ }\href
  {https://doi.org/https://doi.org/10.1016/0370-2693(77)90860-7} {\bibfield
  {journal} {\bibinfo  {journal} {Phys. Lett. B}\ }\textbf {\bibinfo {volume}
  {66}},\ \bibinfo {pages} {205} (\bibinfo {year} {1977})}\BibitemShut
  {NoStop}%
\bibitem [{\citenamefont {Otsuka}\ \emph {et~al.}(1978)\citenamefont {Otsuka},
  \citenamefont {Arima},\ and\ \citenamefont {Iachello}}]{OTSUKA19781}%
  \BibitemOpen
  \bibfield  {author} {\bibinfo {author} {\bibfnamefont {T.}~\bibnamefont
  {Otsuka}}, \bibinfo {author} {\bibfnamefont {A.}~\bibnamefont {Arima}},\ and\
  \bibinfo {author} {\bibfnamefont {F.}~\bibnamefont {Iachello}},\ }\href
  {https://doi.org/https://doi.org/10.1016/0375-9474(78)90532-8} {\bibfield
  {journal} {\bibinfo  {journal} {Nucl. Phys. A}\ }\textbf {\bibinfo {volume}
  {309}},\ \bibinfo {pages} {1} (\bibinfo {year} {1978})}\BibitemShut {NoStop}%
\bibitem [{\citenamefont {Iachello}(1984)}]{Iachello}%
  \BibitemOpen
  \bibfield  {author} {\bibinfo {author} {\bibfnamefont {F.}~\bibnamefont
  {Iachello}},\ }\href {https://doi.org/10.1103/PhysRevLett.53.1427} {\bibfield
   {journal} {\bibinfo  {journal} {Phys. Rev. Lett.}\ }\textbf {\bibinfo
  {volume} {53}},\ \bibinfo {pages} {1427} (\bibinfo {year}
  {1984})}\BibitemShut {NoStop}%
\bibitem [{\citenamefont {Pietralla}\ \emph {et~al.}(2008)\citenamefont
  {Pietralla}, \citenamefont {{von Brentano}},\ and\ \citenamefont
  {Lisetskiy}}]{mspietralla}%
  \BibitemOpen
  \bibfield  {author} {\bibinfo {author} {\bibfnamefont {N.}~\bibnamefont
  {Pietralla}}, \bibinfo {author} {\bibfnamefont {P.}~\bibnamefont {{von
  Brentano}}},\ and\ \bibinfo {author} {\bibfnamefont {A.}~\bibnamefont
  {Lisetskiy}},\ }\href
  {https://doi.org/https://doi.org/10.1016/j.ppnp.2007.08.002} {\bibfield
  {journal} {\bibinfo  {journal} {Prog. Part. Nucl. Phys.}\ }\textbf {\bibinfo
  {volume} {60}},\ \bibinfo {pages} {225} (\bibinfo {year} {2008})}\BibitemShut
  {NoStop}%
\bibitem [{\citenamefont {Heyde}\ and\ \citenamefont {Sau}(1986)}]{heydesau}%
  \BibitemOpen
  \bibfield  {author} {\bibinfo {author} {\bibfnamefont {K.}~\bibnamefont
  {Heyde}}\ and\ \bibinfo {author} {\bibfnamefont {J.}~\bibnamefont {Sau}},\
  }\href {https://doi.org/10.1103/PhysRevC.33.1050} {\bibfield  {journal}
  {\bibinfo  {journal} {Phys. Rev. C}\ }\textbf {\bibinfo {volume} {33}},\
  \bibinfo {pages} {1050} (\bibinfo {year} {1986})}\BibitemShut {NoStop}%
\bibitem [{\citenamefont {Werner}\ \emph {et~al.}(2002)\citenamefont {Werner},
  \citenamefont {Belic}, \citenamefont {{von Brentano}}, \citenamefont
  {Fransen}, \citenamefont {Gade}, \citenamefont {{von Garrel}}, \citenamefont
  {Jolie}, \citenamefont {Kneissl}, \citenamefont {Kohstall}, \citenamefont
  {Linnemann}, \citenamefont {Lisetskiy}, \citenamefont {Pietralla},
  \citenamefont {Pitz}, \citenamefont {Scheck}, \citenamefont {Speidel},
  \citenamefont {Stedile},\ and\ \citenamefont {Yates}}]{werner}%
  \BibitemOpen
  \bibfield  {author} {\bibinfo {author} {\bibfnamefont {V.}~\bibnamefont
  {Werner}}, \bibinfo {author} {\bibfnamefont {D.}~\bibnamefont {Belic}},
  \bibinfo {author} {\bibfnamefont {P.}~\bibnamefont {{von Brentano}}},
  \bibinfo {author} {\bibfnamefont {C.}~\bibnamefont {Fransen}}, \bibinfo
  {author} {\bibfnamefont {A.}~\bibnamefont {Gade}}, \bibinfo {author}
  {\bibfnamefont {H.}~\bibnamefont {{von Garrel}}}, \bibinfo {author}
  {\bibfnamefont {J.}~\bibnamefont {Jolie}}, \bibinfo {author} {\bibfnamefont
  {U.}~\bibnamefont {Kneissl}}, \bibinfo {author} {\bibfnamefont
  {C.}~\bibnamefont {Kohstall}}, \bibinfo {author} {\bibfnamefont
  {A.}~\bibnamefont {Linnemann}}, \bibinfo {author} {\bibfnamefont
  {A.}~\bibnamefont {Lisetskiy}}, \bibinfo {author} {\bibfnamefont
  {N.}~\bibnamefont {Pietralla}}, \bibinfo {author} {\bibfnamefont
  {H.}~\bibnamefont {Pitz}}, \bibinfo {author} {\bibfnamefont {M.}~\bibnamefont
  {Scheck}}, \bibinfo {author} {\bibfnamefont {K.-H.}\ \bibnamefont {Speidel}},
  \bibinfo {author} {\bibfnamefont {F.}~\bibnamefont {Stedile}},\ and\ \bibinfo
  {author} {\bibfnamefont {S.}~\bibnamefont {Yates}},\ }\href
  {https://doi.org/https://doi.org/10.1016/S0370-2693(02)02961-1} {\bibfield
  {journal} {\bibinfo  {journal} {Phys. Lett. B}\ }\textbf {\bibinfo {volume}
  {550}},\ \bibinfo {pages} {140} (\bibinfo {year} {2002})}\BibitemShut
  {NoStop}%
\bibitem [{\citenamefont {Holt}\ \emph {et~al.}(2007)\citenamefont {Holt},
  \citenamefont {Pietralla}, \citenamefont {Holt}, \citenamefont {Kuo},\ and\
  \citenamefont {Rainovski}}]{holt}%
  \BibitemOpen
  \bibfield  {author} {\bibinfo {author} {\bibfnamefont {J.~D.}\ \bibnamefont
  {Holt}}, \bibinfo {author} {\bibfnamefont {N.}~\bibnamefont {Pietralla}},
  \bibinfo {author} {\bibfnamefont {J.~W.}\ \bibnamefont {Holt}}, \bibinfo
  {author} {\bibfnamefont {T.~T.~S.}\ \bibnamefont {Kuo}},\ and\ \bibinfo
  {author} {\bibfnamefont {G.}~\bibnamefont {Rainovski}},\ }\href
  {https://doi.org/10.1103/PhysRevC.76.034325} {\bibfield  {journal} {\bibinfo
  {journal} {Phys. Rev. C}\ }\textbf {\bibinfo {volume} {76}},\ \bibinfo
  {pages} {034325} (\bibinfo {year} {2007})}\BibitemShut {NoStop}%
\bibitem [{\citenamefont {Covelle}\ \emph {et~al.}(2007)\citenamefont
  {Covelle}, \citenamefont {Coraggio}, \citenamefont {Gargano},\ and\
  \citenamefont {Itaco}}]{covello}%
  \BibitemOpen
  \bibfield  {author} {\bibinfo {author} {\bibfnamefont {A.}~\bibnamefont
  {Covelle}}, \bibinfo {author} {\bibfnamefont {L.}~\bibnamefont {Coraggio}},
  \bibinfo {author} {\bibfnamefont {A.}~\bibnamefont {Gargano}},\ and\ \bibinfo
  {author} {\bibfnamefont {N.}~\bibnamefont {Itaco}},\ }\href
  {https://doi.org/https://doi.org/10.1016/j.ppnp.2007.01.001} {\bibfield
  {journal} {\bibinfo  {journal} {Prog. Part. Nucl. Phys.}\ }\textbf {\bibinfo
  {volume} {59}},\ \bibinfo {pages} {401} (\bibinfo {year} {2007})}\BibitemShut
  {NoStop}%
\bibitem [{\citenamefont {Stuchbery}\ and\ \citenamefont {Wood}(2022)}]{stuch}%
  \BibitemOpen
  \bibfield  {author} {\bibinfo {author} {\bibfnamefont {A.~E.}\ \bibnamefont
  {Stuchbery}}\ and\ \bibinfo {author} {\bibfnamefont {J.~L.}\ \bibnamefont
  {Wood}},\ }\bibfield  {title} {\bibinfo {title} {To shell model, or not to
  shell model, that is the question},\ }\href
  {https://doi.org/10.3390/physics4030048} {\bibfield  {journal} {\bibinfo
  {journal} {Physics}\ }\textbf {\bibinfo {volume} {4}},\ \bibinfo {pages}
  {697} (\bibinfo {year} {2022})}\BibitemShut {NoStop}%
\bibitem [{\citenamefont {Hamilton}\ \emph {et~al.}(1984)\citenamefont
  {Hamilton}, \citenamefont {Irb\"ack},\ and\ \citenamefont {Elliott}}]{ham84}%
  \BibitemOpen
  \bibfield  {author} {\bibinfo {author} {\bibfnamefont {W.~D.}\ \bibnamefont
  {Hamilton}}, \bibinfo {author} {\bibfnamefont {A.}~\bibnamefont {Irb\"ack}},\
  and\ \bibinfo {author} {\bibfnamefont {J.~P.}\ \bibnamefont {Elliott}},\
  }\href {https://doi.org/10.1103/PhysRevLett.53.2469} {\bibfield  {journal}
  {\bibinfo  {journal} {Phys. Rev. Lett.}\ }\textbf {\bibinfo {volume} {53}},\
  \bibinfo {pages} {2469} (\bibinfo {year} {1984})}\BibitemShut {NoStop}%
\bibitem [{\citenamefont {Danchev}\ \emph {et~al.}(2011)\citenamefont
  {Danchev}, \citenamefont {Rainovski}, \citenamefont {Pietralla},
  \citenamefont {Gargano}, \citenamefont {Covello}, \citenamefont {Baktash},
  \citenamefont {Beene}, \citenamefont {Bingham}, \citenamefont
  {Galindo-Uribarri}, \citenamefont {Gladnishki}, \citenamefont {Gross},
  \citenamefont {Ponomarev}, \citenamefont {Radford}, \citenamefont
  {Riedinger}, \citenamefont {Scheck}, \citenamefont {Stuchbery}, \citenamefont
  {Wambach}, \citenamefont {Yu},\ and\ \citenamefont {Zamfir}}]{danchev}%
  \BibitemOpen
  \bibfield  {author} {\bibinfo {author} {\bibfnamefont {M.}~\bibnamefont
  {Danchev}}, \bibinfo {author} {\bibfnamefont {G.}~\bibnamefont {Rainovski}},
  \bibinfo {author} {\bibfnamefont {N.}~\bibnamefont {Pietralla}}, \bibinfo
  {author} {\bibfnamefont {A.}~\bibnamefont {Gargano}}, \bibinfo {author}
  {\bibfnamefont {A.}~\bibnamefont {Covello}}, \bibinfo {author} {\bibfnamefont
  {C.}~\bibnamefont {Baktash}}, \bibinfo {author} {\bibfnamefont {J.~R.}\
  \bibnamefont {Beene}}, \bibinfo {author} {\bibfnamefont {C.~R.}\ \bibnamefont
  {Bingham}}, \bibinfo {author} {\bibfnamefont {A.}~\bibnamefont
  {Galindo-Uribarri}}, \bibinfo {author} {\bibfnamefont {K.~A.}\ \bibnamefont
  {Gladnishki}}, \bibinfo {author} {\bibfnamefont {C.~J.}\ \bibnamefont
  {Gross}}, \bibinfo {author} {\bibfnamefont {V.~Y.}\ \bibnamefont
  {Ponomarev}}, \bibinfo {author} {\bibfnamefont {D.~C.}\ \bibnamefont
  {Radford}}, \bibinfo {author} {\bibfnamefont {L.~L.}\ \bibnamefont
  {Riedinger}}, \bibinfo {author} {\bibfnamefont {M.}~\bibnamefont {Scheck}},
  \bibinfo {author} {\bibfnamefont {A.~E.}\ \bibnamefont {Stuchbery}}, \bibinfo
  {author} {\bibfnamefont {J.}~\bibnamefont {Wambach}}, \bibinfo {author}
  {\bibfnamefont {C.-H.}\ \bibnamefont {Yu}},\ and\ \bibinfo {author}
  {\bibfnamefont {N.~V.}\ \bibnamefont {Zamfir}},\ }\href
  {https://doi.org/10.1103/PhysRevC.84.061306} {\bibfield  {journal} {\bibinfo
  {journal} {Phys. Rev. C}\ }\textbf {\bibinfo {volume} {84}},\ \bibinfo
  {pages} {061306} (\bibinfo {year} {2011})}\BibitemShut {NoStop}%
\bibitem [{\citenamefont {Moschner}\ \emph {et~al.}(2016)\citenamefont
  {Moschner}, \citenamefont {Blazhev}, \citenamefont {Jolie}, \citenamefont
  {Warr}, \citenamefont {Boutachkov}, \citenamefont {Bednarczyk}, \citenamefont
  {Sieja}, \citenamefont {Algora}, \citenamefont {Ameil}, \citenamefont
  {Bentley}, \citenamefont {Brambilla}, \citenamefont {Braun}, \citenamefont
  {Camera}, \citenamefont {Cederk\"all}, \citenamefont {Corsi}, \citenamefont
  {Danchev}, \citenamefont {DiJulio}, \citenamefont {Fahlander}, \citenamefont
  {Gerl}, \citenamefont {Giaz}, \citenamefont {Golubev}, \citenamefont
  {G\'orska}, \citenamefont {Grebosz}, \citenamefont {Habermann}, \citenamefont
  {Hackstein}, \citenamefont {Hoischen}, \citenamefont {Kojouharov},
  \citenamefont {Kurz}, \citenamefont {Mărginean}, \citenamefont {Merch\'an},
  \citenamefont {M\"oller}, \citenamefont {Naqvi}, \citenamefont {Nara~Singh},
  \citenamefont {Nociforo}, \citenamefont {Pietralla}, \citenamefont {Pietri},
  \citenamefont {Podoly\'ak}, \citenamefont {Prochazka}, \citenamefont {Reese},
  \citenamefont {Reiter}, \citenamefont {Rudigier}, \citenamefont {Rudolph},
  \citenamefont {Sava}, \citenamefont {Schaffner}, \citenamefont {Scruton},
  \citenamefont {Taprogge}, \citenamefont {Thomas}, \citenamefont {Weick},
  \citenamefont {Wendt}, \citenamefont {Wieland},\ and\ \citenamefont
  {Wollersheim}}]{moschner}%
  \BibitemOpen
  \bibfield  {author} {\bibinfo {author} {\bibfnamefont {K.}~\bibnamefont
  {Moschner}}, \bibinfo {author} {\bibfnamefont {A.}~\bibnamefont {Blazhev}},
  \bibinfo {author} {\bibfnamefont {J.}~\bibnamefont {Jolie}}, \bibinfo
  {author} {\bibfnamefont {N.}~\bibnamefont {Warr}}, \bibinfo {author}
  {\bibfnamefont {P.}~\bibnamefont {Boutachkov}}, \bibinfo {author}
  {\bibfnamefont {P.}~\bibnamefont {Bednarczyk}}, \bibinfo {author}
  {\bibfnamefont {K.}~\bibnamefont {Sieja}}, \bibinfo {author} {\bibfnamefont
  {A.}~\bibnamefont {Algora}}, \bibinfo {author} {\bibfnamefont
  {F.}~\bibnamefont {Ameil}}, \bibinfo {author} {\bibfnamefont {M.~A.}\
  \bibnamefont {Bentley}}, \bibinfo {author} {\bibfnamefont {S.}~\bibnamefont
  {Brambilla}}, \bibinfo {author} {\bibfnamefont {N.}~\bibnamefont {Braun}},
  \bibinfo {author} {\bibfnamefont {F.}~\bibnamefont {Camera}}, \bibinfo
  {author} {\bibfnamefont {J.}~\bibnamefont {Cederk\"all}}, \bibinfo {author}
  {\bibfnamefont {A.}~\bibnamefont {Corsi}}, \bibinfo {author} {\bibfnamefont
  {M.}~\bibnamefont {Danchev}}, \bibinfo {author} {\bibfnamefont
  {D.}~\bibnamefont {DiJulio}}, \bibinfo {author} {\bibfnamefont
  {C.}~\bibnamefont {Fahlander}}, \bibinfo {author} {\bibfnamefont
  {J.}~\bibnamefont {Gerl}}, \bibinfo {author} {\bibfnamefont {A.}~\bibnamefont
  {Giaz}}, \bibinfo {author} {\bibfnamefont {P.}~\bibnamefont {Golubev}},
  \bibinfo {author} {\bibfnamefont {M.}~\bibnamefont {G\'orska}}, \bibinfo
  {author} {\bibfnamefont {J.}~\bibnamefont {Grebosz}}, \bibinfo {author}
  {\bibfnamefont {T.}~\bibnamefont {Habermann}}, \bibinfo {author}
  {\bibfnamefont {M.}~\bibnamefont {Hackstein}}, \bibinfo {author}
  {\bibfnamefont {R.}~\bibnamefont {Hoischen}}, \bibinfo {author}
  {\bibfnamefont {I.}~\bibnamefont {Kojouharov}}, \bibinfo {author}
  {\bibfnamefont {N.}~\bibnamefont {Kurz}}, \bibinfo {author} {\bibfnamefont
  {N.}~\bibnamefont {Mărginean}}, \bibinfo {author} {\bibfnamefont
  {E.}~\bibnamefont {Merch\'an}}, \bibinfo {author} {\bibfnamefont
  {T.}~\bibnamefont {M\"oller}}, \bibinfo {author} {\bibfnamefont
  {F.}~\bibnamefont {Naqvi}}, \bibinfo {author} {\bibfnamefont {B.~S.}\
  \bibnamefont {Nara~Singh}}, \bibinfo {author} {\bibfnamefont
  {C.}~\bibnamefont {Nociforo}}, \bibinfo {author} {\bibfnamefont
  {N.}~\bibnamefont {Pietralla}}, \bibinfo {author} {\bibfnamefont
  {S.}~\bibnamefont {Pietri}}, \bibinfo {author} {\bibfnamefont {{\relax
  Zs.}.}~\bibnamefont {Podoly\'ak}}, \bibinfo {author} {\bibfnamefont
  {A.}~\bibnamefont {Prochazka}}, \bibinfo {author} {\bibfnamefont
  {M.}~\bibnamefont {Reese}}, \bibinfo {author} {\bibfnamefont
  {P.}~\bibnamefont {Reiter}}, \bibinfo {author} {\bibfnamefont
  {M.}~\bibnamefont {Rudigier}}, \bibinfo {author} {\bibfnamefont
  {D.}~\bibnamefont {Rudolph}}, \bibinfo {author} {\bibfnamefont
  {T.}~\bibnamefont {Sava}}, \bibinfo {author} {\bibfnamefont {H.}~\bibnamefont
  {Schaffner}}, \bibinfo {author} {\bibfnamefont {L.}~\bibnamefont {Scruton}},
  \bibinfo {author} {\bibfnamefont {J.}~\bibnamefont {Taprogge}}, \bibinfo
  {author} {\bibfnamefont {T.}~\bibnamefont {Thomas}}, \bibinfo {author}
  {\bibfnamefont {H.}~\bibnamefont {Weick}}, \bibinfo {author} {\bibfnamefont
  {A.}~\bibnamefont {Wendt}}, \bibinfo {author} {\bibfnamefont
  {O.}~\bibnamefont {Wieland}},\ and\ \bibinfo {author} {\bibfnamefont {H.-J.}\
  \bibnamefont {Wollersheim}},\ }\href
  {https://doi.org/10.1103/PhysRevC.94.054323} {\bibfield  {journal} {\bibinfo
  {journal} {Phys. Rev. C}\ }\textbf {\bibinfo {volume} {94}},\ \bibinfo
  {pages} {054323} (\bibinfo {year} {2016})}\BibitemShut {NoStop}%
\bibitem [{\citenamefont {Kern}\ \emph {et~al.}(2020)\citenamefont {Kern},
  \citenamefont {Zidarova}, \citenamefont {Pietralla}, \citenamefont
  {Rainovski}, \citenamefont {Stegmann}, \citenamefont {Blazhev}, \citenamefont
  {Boukhari}, \citenamefont {Cederk\"all}, \citenamefont {Cubiss},
  \citenamefont {Djongolov}, \citenamefont {Fransen}, \citenamefont {Gaffney},
  \citenamefont {Gladnishki}, \citenamefont {Giannopoulos}, \citenamefont
  {Hess}, \citenamefont {Jolie}, \citenamefont {Karayonchev}, \citenamefont
  {Kaya}, \citenamefont {Keatings}, \citenamefont {Kocheva}, \citenamefont
  {Kr\"oll}, \citenamefont {M\"oller}, \citenamefont {O'Neill}, \citenamefont
  {Pakarinen}, \citenamefont {Reiter}, \citenamefont {Rosiak}, \citenamefont
  {Scheck}, \citenamefont {Snall}, \citenamefont {S\"oderstr\"om},
  \citenamefont {Spagnoletti}, \citenamefont {Stoyanova}, \citenamefont
  {Thiel}, \citenamefont {Vogt}, \citenamefont {Warr}, \citenamefont {Welker},
  \citenamefont {Werner}, \citenamefont {Wiederhold},\ and\ \citenamefont
  {De~Witte}}]{kern}%
  \BibitemOpen
  \bibfield  {author} {\bibinfo {author} {\bibfnamefont {R.}~\bibnamefont
  {Kern}}, \bibinfo {author} {\bibfnamefont {R.}~\bibnamefont {Zidarova}},
  \bibinfo {author} {\bibfnamefont {N.}~\bibnamefont {Pietralla}}, \bibinfo
  {author} {\bibfnamefont {G.}~\bibnamefont {Rainovski}}, \bibinfo {author}
  {\bibfnamefont {R.}~\bibnamefont {Stegmann}}, \bibinfo {author}
  {\bibfnamefont {A.}~\bibnamefont {Blazhev}}, \bibinfo {author} {\bibfnamefont
  {A.}~\bibnamefont {Boukhari}}, \bibinfo {author} {\bibfnamefont
  {J.}~\bibnamefont {Cederk\"all}}, \bibinfo {author} {\bibfnamefont {J.~G.}\
  \bibnamefont {Cubiss}}, \bibinfo {author} {\bibfnamefont {M.}~\bibnamefont
  {Djongolov}}, \bibinfo {author} {\bibfnamefont {C.}~\bibnamefont {Fransen}},
  \bibinfo {author} {\bibfnamefont {L.~P.}\ \bibnamefont {Gaffney}}, \bibinfo
  {author} {\bibfnamefont {K.}~\bibnamefont {Gladnishki}}, \bibinfo {author}
  {\bibfnamefont {E.}~\bibnamefont {Giannopoulos}}, \bibinfo {author}
  {\bibfnamefont {H.}~\bibnamefont {Hess}}, \bibinfo {author} {\bibfnamefont
  {J.}~\bibnamefont {Jolie}}, \bibinfo {author} {\bibfnamefont
  {V.}~\bibnamefont {Karayonchev}}, \bibinfo {author} {\bibfnamefont
  {L.}~\bibnamefont {Kaya}}, \bibinfo {author} {\bibfnamefont {J.~M.}\
  \bibnamefont {Keatings}}, \bibinfo {author} {\bibfnamefont {D.}~\bibnamefont
  {Kocheva}}, \bibinfo {author} {\bibfnamefont {T.}~\bibnamefont {Kr\"oll}},
  \bibinfo {author} {\bibfnamefont {O.}~\bibnamefont {M\"oller}}, \bibinfo
  {author} {\bibfnamefont {G.~G.}\ \bibnamefont {O'Neill}}, \bibinfo {author}
  {\bibfnamefont {J.}~\bibnamefont {Pakarinen}}, \bibinfo {author}
  {\bibfnamefont {P.}~\bibnamefont {Reiter}}, \bibinfo {author} {\bibfnamefont
  {D.}~\bibnamefont {Rosiak}}, \bibinfo {author} {\bibfnamefont
  {M.}~\bibnamefont {Scheck}}, \bibinfo {author} {\bibfnamefont
  {J.}~\bibnamefont {Snall}}, \bibinfo {author} {\bibfnamefont {P.-A.}\
  \bibnamefont {S\"oderstr\"om}}, \bibinfo {author} {\bibfnamefont
  {P.}~\bibnamefont {Spagnoletti}}, \bibinfo {author} {\bibfnamefont
  {M.}~\bibnamefont {Stoyanova}}, \bibinfo {author} {\bibfnamefont
  {S.}~\bibnamefont {Thiel}}, \bibinfo {author} {\bibfnamefont
  {A.}~\bibnamefont {Vogt}}, \bibinfo {author} {\bibfnamefont {N.}~\bibnamefont
  {Warr}}, \bibinfo {author} {\bibfnamefont {A.}~\bibnamefont {Welker}},
  \bibinfo {author} {\bibfnamefont {V.}~\bibnamefont {Werner}}, \bibinfo
  {author} {\bibfnamefont {J.}~\bibnamefont {Wiederhold}},\ and\ \bibinfo
  {author} {\bibfnamefont {H.}~\bibnamefont {De~Witte}},\ }\href
  {https://doi.org/10.1103/PhysRevC.102.041304} {\bibfield  {journal} {\bibinfo
   {journal} {Phys. Rev. C}\ }\textbf {\bibinfo {volume} {102}},\ \bibinfo
  {pages} {041304} (\bibinfo {year} {2020})}\BibitemShut {NoStop}%
\bibitem [{\citenamefont {Fransen}\ \emph {et~al.}(2005)\citenamefont
  {Fransen}, \citenamefont {Werner}, \citenamefont {Bandyopadhyay},
  \citenamefont {Boukharouba}, \citenamefont {Lesher}, \citenamefont
  {McEllistrem}, \citenamefont {Jolie}, \citenamefont {Pietralla},
  \citenamefont {Brentano},\ and\ \citenamefont {Yates}}]{zr}%
  \BibitemOpen
  \bibfield  {author} {\bibinfo {author} {\bibfnamefont {C.}~\bibnamefont
  {Fransen}}, \bibinfo {author} {\bibfnamefont {V.}~\bibnamefont {Werner}},
  \bibinfo {author} {\bibfnamefont {D.}~\bibnamefont {Bandyopadhyay}}, \bibinfo
  {author} {\bibfnamefont {N.}~\bibnamefont {Boukharouba}}, \bibinfo {author}
  {\bibfnamefont {S.~R.}\ \bibnamefont {Lesher}}, \bibinfo {author}
  {\bibfnamefont {M.~T.}\ \bibnamefont {McEllistrem}}, \bibinfo {author}
  {\bibfnamefont {J.}~\bibnamefont {Jolie}}, \bibinfo {author} {\bibfnamefont
  {N.}~\bibnamefont {Pietralla}}, \bibinfo {author} {\bibfnamefont {P.~v.}\
  \bibnamefont {Brentano}},\ and\ \bibinfo {author} {\bibfnamefont {S.~W.}\
  \bibnamefont {Yates}},\ }\href {https://doi.org/10.1103/PhysRevC.71.054304}
  {\bibfield  {journal} {\bibinfo  {journal} {Phys. Rev. C}\ }\textbf {\bibinfo
  {volume} {71}},\ \bibinfo {pages} {054304} (\bibinfo {year}
  {2005})}\BibitemShut {NoStop}%
\bibitem [{\citenamefont {Werner}\ \emph {et~al.}(2008)\citenamefont {Werner},
  \citenamefont {Benczer-Koller}, \citenamefont {Kumbartzki}, \citenamefont
  {Holt}, \citenamefont {Boutachkov}, \citenamefont {Stefanova}, \citenamefont
  {Perry}, \citenamefont {Pietralla}, \citenamefont {Ai}, \citenamefont
  {Aleksandrova}, \citenamefont {Anderson}, \citenamefont {Cakirli},
  \citenamefont {Casperson}, \citenamefont {Casten}, \citenamefont
  {Chamberlain}, \citenamefont {Copos}, \citenamefont {Darakchieva},
  \citenamefont {Eckel}, \citenamefont {Evtimova}, \citenamefont {Fitzpatrick},
  \citenamefont {Garnsworthy}, \citenamefont {G\"urdal}, \citenamefont {Heinz},
  \citenamefont {Kovacheva}, \citenamefont {Lambie-Hanson}, \citenamefont
  {Liang}, \citenamefont {Manchev}, \citenamefont {McCutchan}, \citenamefont
  {Meyer}, \citenamefont {Qian}, \citenamefont {Schmidt}, \citenamefont
  {Thompson}, \citenamefont {Williams},\ and\ \citenamefont {Winkler}}]{wex}%
  \BibitemOpen
  \bibfield  {author} {\bibinfo {author} {\bibfnamefont {V.}~\bibnamefont
  {Werner}}, \bibinfo {author} {\bibfnamefont {N.}~\bibnamefont
  {Benczer-Koller}}, \bibinfo {author} {\bibfnamefont {G.}~\bibnamefont
  {Kumbartzki}}, \bibinfo {author} {\bibfnamefont {J.~D.}\ \bibnamefont
  {Holt}}, \bibinfo {author} {\bibfnamefont {P.}~\bibnamefont {Boutachkov}},
  \bibinfo {author} {\bibfnamefont {E.}~\bibnamefont {Stefanova}}, \bibinfo
  {author} {\bibfnamefont {M.}~\bibnamefont {Perry}}, \bibinfo {author}
  {\bibfnamefont {N.}~\bibnamefont {Pietralla}}, \bibinfo {author}
  {\bibfnamefont {H.}~\bibnamefont {Ai}}, \bibinfo {author} {\bibfnamefont
  {K.}~\bibnamefont {Aleksandrova}}, \bibinfo {author} {\bibfnamefont
  {G.}~\bibnamefont {Anderson}}, \bibinfo {author} {\bibfnamefont {R.~B.}\
  \bibnamefont {Cakirli}}, \bibinfo {author} {\bibfnamefont {R.~J.}\
  \bibnamefont {Casperson}}, \bibinfo {author} {\bibfnamefont {R.~F.}\
  \bibnamefont {Casten}}, \bibinfo {author} {\bibfnamefont {M.}~\bibnamefont
  {Chamberlain}}, \bibinfo {author} {\bibfnamefont {C.}~\bibnamefont {Copos}},
  \bibinfo {author} {\bibfnamefont {B.}~\bibnamefont {Darakchieva}}, \bibinfo
  {author} {\bibfnamefont {S.}~\bibnamefont {Eckel}}, \bibinfo {author}
  {\bibfnamefont {M.}~\bibnamefont {Evtimova}}, \bibinfo {author}
  {\bibfnamefont {C.~R.}\ \bibnamefont {Fitzpatrick}}, \bibinfo {author}
  {\bibfnamefont {A.~B.}\ \bibnamefont {Garnsworthy}}, \bibinfo {author}
  {\bibfnamefont {G.}~\bibnamefont {G\"urdal}}, \bibinfo {author}
  {\bibfnamefont {A.}~\bibnamefont {Heinz}}, \bibinfo {author} {\bibfnamefont
  {D.}~\bibnamefont {Kovacheva}}, \bibinfo {author} {\bibfnamefont
  {C.}~\bibnamefont {Lambie-Hanson}}, \bibinfo {author} {\bibfnamefont
  {X.}~\bibnamefont {Liang}}, \bibinfo {author} {\bibfnamefont
  {P.}~\bibnamefont {Manchev}}, \bibinfo {author} {\bibfnamefont {E.~A.}\
  \bibnamefont {McCutchan}}, \bibinfo {author} {\bibfnamefont {D.~A.}\
  \bibnamefont {Meyer}}, \bibinfo {author} {\bibfnamefont {J.}~\bibnamefont
  {Qian}}, \bibinfo {author} {\bibfnamefont {A.}~\bibnamefont {Schmidt}},
  \bibinfo {author} {\bibfnamefont {N.~J.}\ \bibnamefont {Thompson}}, \bibinfo
  {author} {\bibfnamefont {E.}~\bibnamefont {Williams}},\ and\ \bibinfo
  {author} {\bibfnamefont {R.}~\bibnamefont {Winkler}},\ }\href
  {https://doi.org/10.1103/PhysRevC.78.031301} {\bibfield  {journal} {\bibinfo
  {journal} {Phys. Rev. C}\ }\textbf {\bibinfo {volume} {78}},\ \bibinfo
  {pages} {031301} (\bibinfo {year} {2008})}\BibitemShut {NoStop}%
\bibitem [{\citenamefont {Casperson}\ \emph {et~al.}(2013)\citenamefont
  {Casperson}, \citenamefont {Werner},\ and\ \citenamefont {Heinze}}]{casp}%
  \BibitemOpen
  \bibfield  {author} {\bibinfo {author} {\bibfnamefont {R.}~\bibnamefont
  {Casperson}}, \bibinfo {author} {\bibfnamefont {V.}~\bibnamefont {Werner}},\
  and\ \bibinfo {author} {\bibfnamefont {S.}~\bibnamefont {Heinze}},\ }\href
  {https://doi.org/https://doi.org/10.1016/j.physletb.2013.02.042} {\bibfield
  {journal} {\bibinfo  {journal} {Physics Letters B}\ }\textbf {\bibinfo
  {volume} {721}},\ \bibinfo {pages} {51} (\bibinfo {year} {2013})}\BibitemShut
  {NoStop}%
\bibitem [{\citenamefont {Rainovski}\ \emph {et~al.}(2006)\citenamefont
  {Rainovski}, \citenamefont {Pietralla}, \citenamefont {Ahn}, \citenamefont
  {Lister}, \citenamefont {Janssens}, \citenamefont {Carpenter}, \citenamefont
  {Zhu},\ and\ \citenamefont {Barton}}]{rain}%
  \BibitemOpen
  \bibfield  {author} {\bibinfo {author} {\bibfnamefont {G.}~\bibnamefont
  {Rainovski}}, \bibinfo {author} {\bibfnamefont {N.}~\bibnamefont
  {Pietralla}}, \bibinfo {author} {\bibfnamefont {T.}~\bibnamefont {Ahn}},
  \bibinfo {author} {\bibfnamefont {C.~J.}\ \bibnamefont {Lister}}, \bibinfo
  {author} {\bibfnamefont {R.~V.~F.}\ \bibnamefont {Janssens}}, \bibinfo
  {author} {\bibfnamefont {M.~P.}\ \bibnamefont {Carpenter}}, \bibinfo {author}
  {\bibfnamefont {S.}~\bibnamefont {Zhu}},\ and\ \bibinfo {author}
  {\bibfnamefont {C.~J.}\ \bibnamefont {Barton}},\ }\href
  {https://doi.org/10.1103/PhysRevLett.96.122501} {\bibfield  {journal}
  {\bibinfo  {journal} {Phys. Rev. Lett.}\ }\textbf {\bibinfo {volume} {96}},\
  \bibinfo {pages} {122501} (\bibinfo {year} {2006})}\BibitemShut {NoStop}%
\bibitem [{\citenamefont {Stone}\ \emph {et~al.}(2005)\citenamefont {Stone},
  \citenamefont {Stuchbery}, \citenamefont {Danchev}, \citenamefont {Pavan},
  \citenamefont {Timlin}, \citenamefont {Baktash}, \citenamefont {Barton},
  \citenamefont {Beene}, \citenamefont {Benczer-Koller}, \citenamefont
  {Bingham}, \citenamefont {Dupak}, \citenamefont {Galindo-Uribarri},
  \citenamefont {Gross}, \citenamefont {Kumbartzki}, \citenamefont {Radford},
  \citenamefont {Stone},\ and\ \citenamefont {Zamfir}}]{magmom1}%
  \BibitemOpen
  \bibfield  {author} {\bibinfo {author} {\bibfnamefont {N.~J.}\ \bibnamefont
  {Stone}}, \bibinfo {author} {\bibfnamefont {A.~E.}\ \bibnamefont
  {Stuchbery}}, \bibinfo {author} {\bibfnamefont {M.}~\bibnamefont {Danchev}},
  \bibinfo {author} {\bibfnamefont {J.}~\bibnamefont {Pavan}}, \bibinfo
  {author} {\bibfnamefont {C.~L.}\ \bibnamefont {Timlin}}, \bibinfo {author}
  {\bibfnamefont {C.}~\bibnamefont {Baktash}}, \bibinfo {author} {\bibfnamefont
  {C.}~\bibnamefont {Barton}}, \bibinfo {author} {\bibfnamefont
  {J.}~\bibnamefont {Beene}}, \bibinfo {author} {\bibfnamefont
  {N.}~\bibnamefont {Benczer-Koller}}, \bibinfo {author} {\bibfnamefont
  {C.~R.}\ \bibnamefont {Bingham}}, \bibinfo {author} {\bibfnamefont
  {J.}~\bibnamefont {Dupak}}, \bibinfo {author} {\bibfnamefont
  {A.}~\bibnamefont {Galindo-Uribarri}}, \bibinfo {author} {\bibfnamefont
  {C.~J.}\ \bibnamefont {Gross}}, \bibinfo {author} {\bibfnamefont
  {G.}~\bibnamefont {Kumbartzki}}, \bibinfo {author} {\bibfnamefont {D.~C.}\
  \bibnamefont {Radford}}, \bibinfo {author} {\bibfnamefont {J.~R.}\
  \bibnamefont {Stone}},\ and\ \bibinfo {author} {\bibfnamefont {N.~V.}\
  \bibnamefont {Zamfir}},\ }\href
  {https://doi.org/10.1103/PhysRevLett.94.192501} {\bibfield  {journal}
  {\bibinfo  {journal} {Phys. Rev. Lett.}\ }\textbf {\bibinfo {volume} {94}},\
  \bibinfo {pages} {192501} (\bibinfo {year} {2005})}\BibitemShut {NoStop}%
\bibitem [{\citenamefont {Stuchbery}\ and\ \citenamefont
  {Stone}(2007)}]{magmom2}%
  \BibitemOpen
  \bibfield  {author} {\bibinfo {author} {\bibfnamefont {A.~E.}\ \bibnamefont
  {Stuchbery}}\ and\ \bibinfo {author} {\bibfnamefont {N.~J.}\ \bibnamefont
  {Stone}},\ }\href {https://doi.org/10.1103/PhysRevC.76.034307} {\bibfield
  {journal} {\bibinfo  {journal} {Phys. Rev. C}\ }\textbf {\bibinfo {volume}
  {76}},\ \bibinfo {pages} {034307} (\bibinfo {year} {2007})}\BibitemShut
  {NoStop}%
\bibitem [{\citenamefont {Benczer-Koller}\ \emph {et~al.}(2008)\citenamefont
  {Benczer-Koller}, \citenamefont {Kumbartzki}, \citenamefont {Gürdal},
  \citenamefont {Gross}, \citenamefont {Stuchbery}, \citenamefont {Krieger},
  \citenamefont {Hatarik}, \citenamefont {O'Malley}, \citenamefont {Pain},
  \citenamefont {Segen}, \citenamefont {Baktash}, \citenamefont {Beene},
  \citenamefont {Radford}, \citenamefont {Yu}, \citenamefont {Stone},
  \citenamefont {Stone}, \citenamefont {Bingham}, \citenamefont {Danchev},
  \citenamefont {Grzywacz},\ and\ \citenamefont {Mazzocchi}}]{magmom3}%
  \BibitemOpen
  \bibfield  {author} {\bibinfo {author} {\bibfnamefont {N.}~\bibnamefont
  {Benczer-Koller}}, \bibinfo {author} {\bibfnamefont {G.}~\bibnamefont
  {Kumbartzki}}, \bibinfo {author} {\bibfnamefont {G.}~\bibnamefont {Gürdal}},
  \bibinfo {author} {\bibfnamefont {C.}~\bibnamefont {Gross}}, \bibinfo
  {author} {\bibfnamefont {A.}~\bibnamefont {Stuchbery}}, \bibinfo {author}
  {\bibfnamefont {B.}~\bibnamefont {Krieger}}, \bibinfo {author} {\bibfnamefont
  {R.}~\bibnamefont {Hatarik}}, \bibinfo {author} {\bibfnamefont
  {P.}~\bibnamefont {O'Malley}}, \bibinfo {author} {\bibfnamefont
  {S.}~\bibnamefont {Pain}}, \bibinfo {author} {\bibfnamefont {L.}~\bibnamefont
  {Segen}}, \bibinfo {author} {\bibfnamefont {C.}~\bibnamefont {Baktash}},
  \bibinfo {author} {\bibfnamefont {J.}~\bibnamefont {Beene}}, \bibinfo
  {author} {\bibfnamefont {D.}~\bibnamefont {Radford}}, \bibinfo {author}
  {\bibfnamefont {C.}~\bibnamefont {Yu}}, \bibinfo {author} {\bibfnamefont
  {N.}~\bibnamefont {Stone}}, \bibinfo {author} {\bibfnamefont
  {J.}~\bibnamefont {Stone}}, \bibinfo {author} {\bibfnamefont
  {C.}~\bibnamefont {Bingham}}, \bibinfo {author} {\bibfnamefont
  {M.}~\bibnamefont {Danchev}}, \bibinfo {author} {\bibfnamefont
  {R.}~\bibnamefont {Grzywacz}},\ and\ \bibinfo {author} {\bibfnamefont
  {C.}~\bibnamefont {Mazzocchi}},\ }\href
  {https://doi.org/https://doi.org/10.1016/j.physletb.2008.05.048} {\bibfield
  {journal} {\bibinfo  {journal} {Phys. Lett. B}\ }\textbf {\bibinfo {volume}
  {664}},\ \bibinfo {pages} {241} (\bibinfo {year} {2008})}\BibitemShut
  {NoStop}%
\bibitem [{\citenamefont {Hughes}\ \emph {et~al.}(2005)\citenamefont {Hughes},
  \citenamefont {Zamfir}, \citenamefont {Radford}, \citenamefont {Gross},
  \citenamefont {Barton}, \citenamefont {Baktash}, \citenamefont {Caprio},
  \citenamefont {Casten}, \citenamefont {Galindo-Uribarri}, \citenamefont
  {Hausladen}, \citenamefont {McCutchan}, \citenamefont {Ressler},
  \citenamefont {Shapira}, \citenamefont {Stracener},\ and\ \citenamefont
  {Yu}}]{hughes}%
  \BibitemOpen
  \bibfield  {author} {\bibinfo {author} {\bibfnamefont {R.~O.}\ \bibnamefont
  {Hughes}}, \bibinfo {author} {\bibfnamefont {N.~V.}\ \bibnamefont {Zamfir}},
  \bibinfo {author} {\bibfnamefont {D.~C.}\ \bibnamefont {Radford}}, \bibinfo
  {author} {\bibfnamefont {C.~J.}\ \bibnamefont {Gross}}, \bibinfo {author}
  {\bibfnamefont {C.~J.}\ \bibnamefont {Barton}}, \bibinfo {author}
  {\bibfnamefont {C.}~\bibnamefont {Baktash}}, \bibinfo {author} {\bibfnamefont
  {M.~A.}\ \bibnamefont {Caprio}}, \bibinfo {author} {\bibfnamefont {R.~F.}\
  \bibnamefont {Casten}}, \bibinfo {author} {\bibfnamefont {A.}~\bibnamefont
  {Galindo-Uribarri}}, \bibinfo {author} {\bibfnamefont {P.~A.}\ \bibnamefont
  {Hausladen}}, \bibinfo {author} {\bibfnamefont {E.~A.}\ \bibnamefont
  {McCutchan}}, \bibinfo {author} {\bibfnamefont {J.~J.}\ \bibnamefont
  {Ressler}}, \bibinfo {author} {\bibfnamefont {D.}~\bibnamefont {Shapira}},
  \bibinfo {author} {\bibfnamefont {D.~W.}\ \bibnamefont {Stracener}},\ and\
  \bibinfo {author} {\bibfnamefont {C.-H.}\ \bibnamefont {Yu}},\ }\href
  {https://doi.org/10.1103/PhysRevC.71.044311} {\bibfield  {journal} {\bibinfo
  {journal} {Phys. Rev. C}\ }\textbf {\bibinfo {volume} {71}},\ \bibinfo
  {pages} {044311} (\bibinfo {year} {2005})}\BibitemShut {NoStop}%
\bibitem [{\citenamefont {Bucurescu}\ \emph {et~al.}(2016)\citenamefont
  {Bucurescu}, \citenamefont {Căta-Danil}, \citenamefont {Ciocan},
  \citenamefont {Costache}, \citenamefont {Deleanu}, \citenamefont {Dima},
  \citenamefont {Filipescu}, \citenamefont {Florea}, \citenamefont {Ghiţă},
  \citenamefont {Glodariu}, \citenamefont {Ivaşcu}, \citenamefont {Lică},
  \citenamefont {Mărginean}, \citenamefont {Mărginean}, \citenamefont
  {Mihai}, \citenamefont {Negret}, \citenamefont {Niţă}, \citenamefont
  {Olăcel}, \citenamefont {Pascu}, \citenamefont {Sava}, \citenamefont
  {Stroe}, \citenamefont {Şerban}, \citenamefont {Şuvăilă}, \citenamefont
  {Toma}, \citenamefont {Zamfir}, \citenamefont {Căta-Danil}, \citenamefont
  {Gheorghe}, \citenamefont {Mitu}, \citenamefont {Suliman}, \citenamefont
  {Ur}, \citenamefont {Braunroth}, \citenamefont {Dewald}, \citenamefont
  {Fransen}, \citenamefont {Bruce}, \citenamefont {Podolyák}, \citenamefont
  {Regan},\ and\ \citenamefont {Roberts}}]{rosphere}%
  \BibitemOpen
  \bibfield  {author} {\bibinfo {author} {\bibfnamefont {D.}~\bibnamefont
  {Bucurescu}}, \bibinfo {author} {\bibfnamefont {I.}~\bibnamefont
  {Căta-Danil}}, \bibinfo {author} {\bibfnamefont {G.}~\bibnamefont {Ciocan}},
  \bibinfo {author} {\bibfnamefont {C.}~\bibnamefont {Costache}}, \bibinfo
  {author} {\bibfnamefont {D.}~\bibnamefont {Deleanu}}, \bibinfo {author}
  {\bibfnamefont {R.}~\bibnamefont {Dima}}, \bibinfo {author} {\bibfnamefont
  {D.}~\bibnamefont {Filipescu}}, \bibinfo {author} {\bibfnamefont
  {N.}~\bibnamefont {Florea}}, \bibinfo {author} {\bibfnamefont
  {D.}~\bibnamefont {Ghiţă}}, \bibinfo {author} {\bibfnamefont
  {T.}~\bibnamefont {Glodariu}}, \bibinfo {author} {\bibfnamefont
  {M.}~\bibnamefont {Ivaşcu}}, \bibinfo {author} {\bibfnamefont
  {R.}~\bibnamefont {Lică}}, \bibinfo {author} {\bibfnamefont
  {N.}~\bibnamefont {Mărginean}}, \bibinfo {author} {\bibfnamefont
  {R.}~\bibnamefont {Mărginean}}, \bibinfo {author} {\bibfnamefont
  {C.}~\bibnamefont {Mihai}}, \bibinfo {author} {\bibfnamefont
  {A.}~\bibnamefont {Negret}}, \bibinfo {author} {\bibfnamefont
  {C.}~\bibnamefont {Niţă}}, \bibinfo {author} {\bibfnamefont
  {A.}~\bibnamefont {Olăcel}}, \bibinfo {author} {\bibfnamefont
  {S.}~\bibnamefont {Pascu}}, \bibinfo {author} {\bibfnamefont
  {T.}~\bibnamefont {Sava}}, \bibinfo {author} {\bibfnamefont {L.}~\bibnamefont
  {Stroe}}, \bibinfo {author} {\bibfnamefont {A.}~\bibnamefont {Şerban}},
  \bibinfo {author} {\bibfnamefont {R.}~\bibnamefont {Şuvăilă}}, \bibinfo
  {author} {\bibfnamefont {S.}~\bibnamefont {Toma}}, \bibinfo {author}
  {\bibfnamefont {N.}~\bibnamefont {Zamfir}}, \bibinfo {author} {\bibfnamefont
  {G.}~\bibnamefont {Căta-Danil}}, \bibinfo {author} {\bibfnamefont
  {I.}~\bibnamefont {Gheorghe}}, \bibinfo {author} {\bibfnamefont
  {I.}~\bibnamefont {Mitu}}, \bibinfo {author} {\bibfnamefont {G.}~\bibnamefont
  {Suliman}}, \bibinfo {author} {\bibfnamefont {C.}~\bibnamefont {Ur}},
  \bibinfo {author} {\bibfnamefont {T.}~\bibnamefont {Braunroth}}, \bibinfo
  {author} {\bibfnamefont {A.}~\bibnamefont {Dewald}}, \bibinfo {author}
  {\bibfnamefont {C.}~\bibnamefont {Fransen}}, \bibinfo {author} {\bibfnamefont
  {A.}~\bibnamefont {Bruce}}, \bibinfo {author} {\bibfnamefont
  {Z.}~\bibnamefont {Podolyák}}, \bibinfo {author} {\bibfnamefont
  {P.}~\bibnamefont {Regan}},\ and\ \bibinfo {author} {\bibfnamefont
  {O.}~\bibnamefont {Roberts}},\ }\href
  {https://doi.org/https://doi.org/10.1016/j.nima.2016.08.052} {\bibfield
  {journal} {\bibinfo  {journal} {NIM-A}\ }\textbf {\bibinfo {volume} {837}},\
  \bibinfo {pages} {1} (\bibinfo {year} {2016})}\BibitemShut {NoStop}%
\bibitem [{\citenamefont {Beck}\ \emph {et~al.}(2020)\citenamefont {Beck},
  \citenamefont {Costache}, \citenamefont {Lică}, \citenamefont {Mărginean},
  \citenamefont {Mihai}, \citenamefont {Mihai}, \citenamefont {Papst},
  \citenamefont {Pascu}, \citenamefont {Pietralla}, \citenamefont {Sotty},
  \citenamefont {Stan}, \citenamefont {Turturică}, \citenamefont {Werner},
  \citenamefont {Wiederhold},\ and\ \citenamefont {Witt}}]{sorcerer}%
  \BibitemOpen
  \bibfield  {author} {\bibinfo {author} {\bibfnamefont {T.}~\bibnamefont
  {Beck}}, \bibinfo {author} {\bibfnamefont {C.}~\bibnamefont {Costache}},
  \bibinfo {author} {\bibfnamefont {R.}~\bibnamefont {Lică}}, \bibinfo
  {author} {\bibfnamefont {N.}~\bibnamefont {Mărginean}}, \bibinfo {author}
  {\bibfnamefont {C.}~\bibnamefont {Mihai}}, \bibinfo {author} {\bibfnamefont
  {R.}~\bibnamefont {Mihai}}, \bibinfo {author} {\bibfnamefont
  {O.}~\bibnamefont {Papst}}, \bibinfo {author} {\bibfnamefont
  {S.}~\bibnamefont {Pascu}}, \bibinfo {author} {\bibfnamefont
  {N.}~\bibnamefont {Pietralla}}, \bibinfo {author} {\bibfnamefont
  {C.}~\bibnamefont {Sotty}}, \bibinfo {author} {\bibfnamefont
  {L.}~\bibnamefont {Stan}}, \bibinfo {author} {\bibfnamefont {A.}~\bibnamefont
  {Turturică}}, \bibinfo {author} {\bibfnamefont {V.}~\bibnamefont {Werner}},
  \bibinfo {author} {\bibfnamefont {J.}~\bibnamefont {Wiederhold}},\ and\
  \bibinfo {author} {\bibfnamefont {W.}~\bibnamefont {Witt}},\ }\href
  {https://doi.org/https://doi.org/10.1016/j.nima.2019.163090} {\bibfield
  {journal} {\bibinfo  {journal} {NIM-A}\ }\textbf {\bibinfo {volume} {951}},\
  \bibinfo {pages} {163090} (\bibinfo {year} {2020})}\BibitemShut {NoStop}%
\bibitem [{\citenamefont {Stahl}\ \emph {et~al.}(2017)\citenamefont {Stahl},
  \citenamefont {Leske}, \citenamefont {Lettmann},\ and\ \citenamefont
  {Pietralla}}]{apcad}%
  \BibitemOpen
  \bibfield  {author} {\bibinfo {author} {\bibfnamefont {C.}~\bibnamefont
  {Stahl}}, \bibinfo {author} {\bibfnamefont {J.}~\bibnamefont {Leske}},
  \bibinfo {author} {\bibfnamefont {M.}~\bibnamefont {Lettmann}},\ and\
  \bibinfo {author} {\bibfnamefont {N.}~\bibnamefont {Pietralla}},\ }\href
  {https://doi.org/https://doi.org/10.1016/j.cpc.2017.01.009} {\bibfield
  {journal} {\bibinfo  {journal} {Comp. Phys. Com.}\ }\textbf {\bibinfo
  {volume} {214}},\ \bibinfo {pages} {174} (\bibinfo {year}
  {2017})}\BibitemShut {NoStop}%
\bibitem [{\citenamefont {Ziegler}\ and\ \citenamefont
  {Biersack}(1985)}]{srim}%
  \BibitemOpen
  \bibfield  {author} {\bibinfo {author} {\bibfnamefont {J.~F.}\ \bibnamefont
  {Ziegler}}\ and\ \bibinfo {author} {\bibfnamefont {J.~P.}\ \bibnamefont
  {Biersack}},\ }in\ \href@noop {} {\emph {\bibinfo {booktitle} {Treatise on
  Heavy-Ion Science}}},\ Vol.~\bibinfo {volume} {6},\ \bibinfo {editor} {edited
  by\ \bibinfo {editor} {\bibfnamefont {D.~A.}\ \bibnamefont {Bromley}}}\
  (\bibinfo  {publisher} {Astrophysics, Chemistry, and Condensed Matter},\
  \bibinfo {address} {New York},\ \bibinfo {year} {1985})\ Chap.~\bibinfo
  {chapter} {3}, pp.\ \bibinfo {pages} {95--129}\BibitemShut {NoStop}%
\bibitem [{\citenamefont {Ziegler}\ \emph {et~al.}(2010)\citenamefont
  {Ziegler}, \citenamefont {Ziegler},\ and\ \citenamefont {Biersack}}]{srim2}%
  \BibitemOpen
  \bibfield  {author} {\bibinfo {author} {\bibfnamefont {J.~F.}\ \bibnamefont
  {Ziegler}}, \bibinfo {author} {\bibfnamefont {M.}~\bibnamefont {Ziegler}},\
  and\ \bibinfo {author} {\bibfnamefont {J.}~\bibnamefont {Biersack}},\ }\href
  {https://doi.org/https://doi.org/10.1016/j.nimb.2010.02.091} {\bibfield
  {journal} {\bibinfo  {journal} {Nuclear Instruments and Methods in Physics
  Research Section B: Beam Interactions with Materials and Atoms}\ }\textbf
  {\bibinfo {volume} {268}},\ \bibinfo {pages} {1818} (\bibinfo {year}
  {2010})},\ \bibinfo {note} {19th International Conference on Ion Beam
  Analysis}\BibitemShut {NoStop}%
\bibitem [{\citenamefont {James}(1994)}]{minuit}%
  \BibitemOpen
  \bibfield  {author} {\bibinfo {author} {\bibfnamefont {F.}~\bibnamefont
  {James}},\ }\href@noop {} {}\bibinfo {type} {CERN Program Library Long
  Writeup D506}\ \bibinfo {number} {Version 94.1}\ (\bibinfo  {institution}
  {CERN},\ \bibinfo {year} {1994})\BibitemShut {NoStop}%
\bibitem [{\citenamefont {Radford}\ \emph {et~al.}(2002)\citenamefont
  {Radford}, \citenamefont {Baktash}, \citenamefont {Beene}, \citenamefont
  {Fuentes}, \citenamefont {Galindo-Uribarri}, \citenamefont {Gross},
  \citenamefont {Hausladen}, \citenamefont {Lewis}, \citenamefont {Mueller},
  \citenamefont {Padilla}, \citenamefont {Shapira}, \citenamefont {Stracener},
  \citenamefont {Yu}, \citenamefont {Barton}, \citenamefont {Caprio},
  \citenamefont {Coraggio}, \citenamefont {Covello}, \citenamefont {Gargano},
  \citenamefont {Hartley},\ and\ \citenamefont {Zamfir}}]{radford}%
  \BibitemOpen
  \bibfield  {author} {\bibinfo {author} {\bibfnamefont {D.~C.}\ \bibnamefont
  {Radford}}, \bibinfo {author} {\bibfnamefont {C.}~\bibnamefont {Baktash}},
  \bibinfo {author} {\bibfnamefont {J.~R.}\ \bibnamefont {Beene}}, \bibinfo
  {author} {\bibfnamefont {B.}~\bibnamefont {Fuentes}}, \bibinfo {author}
  {\bibfnamefont {A.}~\bibnamefont {Galindo-Uribarri}}, \bibinfo {author}
  {\bibfnamefont {C.~J.}\ \bibnamefont {Gross}}, \bibinfo {author}
  {\bibfnamefont {P.~A.}\ \bibnamefont {Hausladen}}, \bibinfo {author}
  {\bibfnamefont {T.~A.}\ \bibnamefont {Lewis}}, \bibinfo {author}
  {\bibfnamefont {P.~E.}\ \bibnamefont {Mueller}}, \bibinfo {author}
  {\bibfnamefont {E.}~\bibnamefont {Padilla}}, \bibinfo {author} {\bibfnamefont
  {D.}~\bibnamefont {Shapira}}, \bibinfo {author} {\bibfnamefont {D.~W.}\
  \bibnamefont {Stracener}}, \bibinfo {author} {\bibfnamefont {C.-H.}\
  \bibnamefont {Yu}}, \bibinfo {author} {\bibfnamefont {C.~J.}\ \bibnamefont
  {Barton}}, \bibinfo {author} {\bibfnamefont {M.~A.}\ \bibnamefont {Caprio}},
  \bibinfo {author} {\bibfnamefont {L.}~\bibnamefont {Coraggio}}, \bibinfo
  {author} {\bibfnamefont {A.}~\bibnamefont {Covello}}, \bibinfo {author}
  {\bibfnamefont {A.}~\bibnamefont {Gargano}}, \bibinfo {author} {\bibfnamefont
  {D.~J.}\ \bibnamefont {Hartley}},\ and\ \bibinfo {author} {\bibfnamefont
  {N.~V.}\ \bibnamefont {Zamfir}},\ }\href
  {https://doi.org/10.1103/PhysRevLett.88.222501} {\bibfield  {journal}
  {\bibinfo  {journal} {Phys. Rev. Lett.}\ }\textbf {\bibinfo {volume} {88}},\
  \bibinfo {pages} {222501} (\bibinfo {year} {2002})}\BibitemShut {NoStop}%
\bibitem [{\citenamefont {Brown}\ \emph {et~al.}(2005)\citenamefont {Brown},
  \citenamefont {Stone}, \citenamefont {Stone}, \citenamefont {Towner},\ and\
  \citenamefont {Hjorth-Jensen}}]{BrownSN100PN}%
  \BibitemOpen
  \bibfield  {author} {\bibinfo {author} {\bibfnamefont {B.~A.}\ \bibnamefont
  {Brown}}, \bibinfo {author} {\bibfnamefont {N.~J.}\ \bibnamefont {Stone}},
  \bibinfo {author} {\bibfnamefont {J.~R.}\ \bibnamefont {Stone}}, \bibinfo
  {author} {\bibfnamefont {I.~S.}\ \bibnamefont {Towner}},\ and\ \bibinfo
  {author} {\bibfnamefont {M.}~\bibnamefont {Hjorth-Jensen}},\ }\href
  {https://doi.org/doi.org/10.1103/PhysRevC.71.044317} {\bibfield  {journal}
  {\bibinfo  {journal} {Phys. Rev. C}\ }\textbf {\bibinfo {volume} {71}},\
  \bibinfo {pages} {044317} (\bibinfo {year} {2005})}\BibitemShut {NoStop}%
\bibitem [{\citenamefont {Khazov}\ \emph {et~al.}(2005)\citenamefont {Khazov},
  \citenamefont {Rodionov}, \citenamefont {Sakharov},\ and\ \citenamefont
  {Singh}}]{nucldata}%
  \BibitemOpen
  \bibfield  {author} {\bibinfo {author} {\bibfnamefont {Y.}~\bibnamefont
  {Khazov}}, \bibinfo {author} {\bibfnamefont {A.}~\bibnamefont {Rodionov}},
  \bibinfo {author} {\bibfnamefont {S.}~\bibnamefont {Sakharov}},\ and\
  \bibinfo {author} {\bibfnamefont {B.}~\bibnamefont {Singh}},\ }\href
  {https://doi.org/https://doi.org/10.1016/j.nds.2005.03.001} {\bibfield
  {journal} {\bibinfo  {journal} {Nucl. Data Sheets}\ }\textbf {\bibinfo
  {volume} {104}},\ \bibinfo {pages} {497} (\bibinfo {year}
  {2005})}\BibitemShut {NoStop}%
\bibitem [{\citenamefont {Shimizu}(2013)}]{kshell1}%
  \BibitemOpen
  \bibfield  {author} {\bibinfo {author} {\bibfnamefont {N.}~\bibnamefont
  {Shimizu}},\ }\href
  {https://doi.org/https://doi.org/10.48550/arXiv.1310.5431} {} (\bibinfo
  {year} {2013}),\ \Eprint {https://arxiv.org/abs/1310.5431} {arXiv:1310.5431
  [nucl-th]} \BibitemShut {NoStop}%
\bibitem [{\citenamefont {Schimizu}\ \emph {et~al.}(2019)\citenamefont
  {Schimizu}, \citenamefont {Mizusaki}, \citenamefont {Utsuno},\ and\
  \citenamefont {Tsunoda}}]{kshell2}%
  \BibitemOpen
  \bibfield  {author} {\bibinfo {author} {\bibfnamefont {N.}~\bibnamefont
  {Schimizu}}, \bibinfo {author} {\bibfnamefont {T.}~\bibnamefont {Mizusaki}},
  \bibinfo {author} {\bibfnamefont {Y.}~\bibnamefont {Utsuno}},\ and\ \bibinfo
  {author} {\bibfnamefont {Y.}~\bibnamefont {Tsunoda}},\ }\href
  {https://doi.org/https://doi.org/10.1016/j.cpc.2019.06.011} {\bibfield
  {journal} {\bibinfo  {journal} {Comp. Phys. Comm.}\ }\textbf {\bibinfo
  {volume} {244}},\ \bibinfo {pages} {372} (\bibinfo {year}
  {2019})}\BibitemShut {NoStop}%
\bibitem [{\citenamefont {Gray}\ \emph {et~al.}(2020)\citenamefont {Gray},
  \citenamefont {Allmond}, \citenamefont {Stuchbery}, \citenamefont {Yu},
  \citenamefont {Baktash}, \citenamefont {Gargano}, \citenamefont
  {Galindo-Uribarri}, \citenamefont {Radford}, \citenamefont {Batchelder},
  \citenamefont {Beene}, \citenamefont {Bingham}, \citenamefont {Coraggio},
  \citenamefont {Covello}, \citenamefont {Danchev}, \citenamefont {Gross},
  \citenamefont {Hausladen}, \citenamefont {Itaco}, \citenamefont {Krolas},
  \citenamefont {Liang}, \citenamefont {Padilla-Rodal}, \citenamefont {Pavan},
  \citenamefont {Stracener},\ and\ \citenamefont {Varner}}]{gray2020}%
  \BibitemOpen
  \bibfield  {author} {\bibinfo {author} {\bibfnamefont {T.~J.}\ \bibnamefont
  {Gray}}, \bibinfo {author} {\bibfnamefont {J.~M.}\ \bibnamefont {Allmond}},
  \bibinfo {author} {\bibfnamefont {A.~E.}\ \bibnamefont {Stuchbery}}, \bibinfo
  {author} {\bibfnamefont {C.-H.}\ \bibnamefont {Yu}}, \bibinfo {author}
  {\bibfnamefont {C.}~\bibnamefont {Baktash}}, \bibinfo {author} {\bibfnamefont
  {A.}~\bibnamefont {Gargano}}, \bibinfo {author} {\bibfnamefont
  {A.}~\bibnamefont {Galindo-Uribarri}}, \bibinfo {author} {\bibfnamefont
  {D.~C.}\ \bibnamefont {Radford}}, \bibinfo {author} {\bibfnamefont {J.~C.}\
  \bibnamefont {Batchelder}}, \bibinfo {author} {\bibfnamefont {J.~R.}\
  \bibnamefont {Beene}}, \bibinfo {author} {\bibfnamefont {C.~R.}\ \bibnamefont
  {Bingham}}, \bibinfo {author} {\bibfnamefont {L.}~\bibnamefont {Coraggio}},
  \bibinfo {author} {\bibfnamefont {A.}~\bibnamefont {Covello}}, \bibinfo
  {author} {\bibfnamefont {M.}~\bibnamefont {Danchev}}, \bibinfo {author}
  {\bibfnamefont {C.~J.}\ \bibnamefont {Gross}}, \bibinfo {author}
  {\bibfnamefont {P.~A.}\ \bibnamefont {Hausladen}}, \bibinfo {author}
  {\bibfnamefont {N.}~\bibnamefont {Itaco}}, \bibinfo {author} {\bibfnamefont
  {W.}~\bibnamefont {Krolas}}, \bibinfo {author} {\bibfnamefont {J.~F.}\
  \bibnamefont {Liang}}, \bibinfo {author} {\bibfnamefont {E.}~\bibnamefont
  {Padilla-Rodal}}, \bibinfo {author} {\bibfnamefont {J.}~\bibnamefont
  {Pavan}}, \bibinfo {author} {\bibfnamefont {D.~W.}\ \bibnamefont
  {Stracener}},\ and\ \bibinfo {author} {\bibfnamefont {R.~L.}\ \bibnamefont
  {Varner}},\ }\href {https://doi.org/doi.org/10.1103/PhysRevLett.124.032502}
  {\bibfield  {journal} {\bibinfo  {journal} {Phys. Rev. Lett.}\ }\textbf
  {\bibinfo {volume} {124}},\ \bibinfo {pages} {032502} (\bibinfo {year}
  {2020})}\BibitemShut {NoStop}%
\bibitem [{\citenamefont {Hicks}\ \emph {et~al.}(2022)\citenamefont {Hicks},
  \citenamefont {Stuchbery}, \citenamefont {Churchill}, \citenamefont
  {Bandyopadhyay}, \citenamefont {R.}, \citenamefont {Coombes},\ and\
  \citenamefont {Davoren}}]{hicks2022}%
  \BibitemOpen
  \bibfield  {author} {\bibinfo {author} {\bibfnamefont {S.~F.}\ \bibnamefont
  {Hicks}}, \bibinfo {author} {\bibfnamefont {A.~E.}\ \bibnamefont
  {Stuchbery}}, \bibinfo {author} {\bibfnamefont {T.~H.}\ \bibnamefont
  {Churchill}}, \bibinfo {author} {\bibfnamefont {D.}~\bibnamefont
  {Bandyopadhyay}}, \bibinfo {author} {\bibfnamefont {C.~B.}\ \bibnamefont
  {R.}}, \bibinfo {author} {\bibfnamefont {B.~J.}\ \bibnamefont {Coombes}},\
  and\ \bibinfo {author} {\bibfnamefont {C.~M.}\ \bibnamefont {Davoren}},\
  }\href {https://doi.org/doi.org/10.1103/PhysRevC.105.024329} {\bibfield
  {journal} {\bibinfo  {journal} {Phys. Rev. C}\ }\textbf {\bibinfo {volume}
  {105}},\ \bibinfo {pages} {024329} (\bibinfo {year} {2022})}\BibitemShut
  {NoStop}%
\bibitem [{\citenamefont {Pietralla}\ \emph {et~al.}(1998)\citenamefont
  {Pietralla}, \citenamefont {Belic}, \citenamefont {von Brentano},
  \citenamefont {Fransen}, \citenamefont {Herzberg}, \citenamefont {Kneissl},
  \citenamefont {Maser}, \citenamefont {Matschinsky}, \citenamefont {Nord},
  \citenamefont {Otsuka}, \citenamefont {Pitz}, \citenamefont {Werner},\ and\
  \citenamefont {Wiedenh\"over}}]{barium}%
  \BibitemOpen
  \bibfield  {author} {\bibinfo {author} {\bibfnamefont {N.}~\bibnamefont
  {Pietralla}}, \bibinfo {author} {\bibfnamefont {D.}~\bibnamefont {Belic}},
  \bibinfo {author} {\bibfnamefont {P.}~\bibnamefont {von Brentano}}, \bibinfo
  {author} {\bibfnamefont {C.}~\bibnamefont {Fransen}}, \bibinfo {author}
  {\bibfnamefont {R.-D.}\ \bibnamefont {Herzberg}}, \bibinfo {author}
  {\bibfnamefont {U.}~\bibnamefont {Kneissl}}, \bibinfo {author} {\bibfnamefont
  {H.}~\bibnamefont {Maser}}, \bibinfo {author} {\bibfnamefont
  {P.}~\bibnamefont {Matschinsky}}, \bibinfo {author} {\bibfnamefont
  {A.}~\bibnamefont {Nord}}, \bibinfo {author} {\bibfnamefont {T.}~\bibnamefont
  {Otsuka}}, \bibinfo {author} {\bibfnamefont {H.~H.}\ \bibnamefont {Pitz}},
  \bibinfo {author} {\bibfnamefont {V.}~\bibnamefont {Werner}},\ and\ \bibinfo
  {author} {\bibfnamefont {I.}~\bibnamefont {Wiedenh\"over}},\ }\href
  {https://doi.org/10.1103/PhysRevC.58.796} {\bibfield  {journal} {\bibinfo
  {journal} {Phys. Rev. C}\ }\textbf {\bibinfo {volume} {58}},\ \bibinfo
  {pages} {796} (\bibinfo {year} {1998})}\BibitemShut {NoStop}%
\bibitem [{\citenamefont {Williams}\ \emph {et~al.}(2009)\citenamefont
  {Williams}, \citenamefont {Casperson}, \citenamefont {Werner}, \citenamefont
  {Ai}, \citenamefont {Boutachkov}, \citenamefont {Chamberlain}, \citenamefont
  {G\"urdal}, \citenamefont {Heinz}, \citenamefont {McCutchan}, \citenamefont
  {Qian},\ and\ \citenamefont {Winkler}}]{williams}%
  \BibitemOpen
  \bibfield  {author} {\bibinfo {author} {\bibfnamefont {E.}~\bibnamefont
  {Williams}}, \bibinfo {author} {\bibfnamefont {R.~J.}\ \bibnamefont
  {Casperson}}, \bibinfo {author} {\bibfnamefont {V.}~\bibnamefont {Werner}},
  \bibinfo {author} {\bibfnamefont {H.}~\bibnamefont {Ai}}, \bibinfo {author}
  {\bibfnamefont {P.}~\bibnamefont {Boutachkov}}, \bibinfo {author}
  {\bibfnamefont {M.}~\bibnamefont {Chamberlain}}, \bibinfo {author}
  {\bibfnamefont {G.}~\bibnamefont {G\"urdal}}, \bibinfo {author}
  {\bibfnamefont {A.}~\bibnamefont {Heinz}}, \bibinfo {author} {\bibfnamefont
  {E.~A.}\ \bibnamefont {McCutchan}}, \bibinfo {author} {\bibfnamefont
  {J.}~\bibnamefont {Qian}},\ and\ \bibinfo {author} {\bibfnamefont
  {R.}~\bibnamefont {Winkler}},\ }\href
  {https://doi.org/10.1103/PhysRevC.80.054309} {\bibfield  {journal} {\bibinfo
  {journal} {Phys. Rev. C}\ }\textbf {\bibinfo {volume} {80}},\ \bibinfo
  {pages} {054309} (\bibinfo {year} {2009})}\BibitemShut {NoStop}%
\bibitem [{\citenamefont {Gladnishki}\ \emph {et~al.}(2010)\citenamefont
  {Gladnishki}, \citenamefont {Rainovski}, \citenamefont {Petkov},
  \citenamefont {Jolie}, \citenamefont {Pietralla}, \citenamefont {Blazhev},
  \citenamefont {Damyanova}, \citenamefont {Danchev}, \citenamefont {Dewald},
  \citenamefont {Fransen}, \citenamefont {Hackstein}, \citenamefont
  {Karagyozov}, \citenamefont {M\"oller}, \citenamefont {Pissulla},
  \citenamefont {Reese}, \citenamefont {Rother},\ and\ \citenamefont
  {Topchiyska}}]{Gladnishki}%
  \BibitemOpen
  \bibfield  {author} {\bibinfo {author} {\bibfnamefont {K.~A.}\ \bibnamefont
  {Gladnishki}}, \bibinfo {author} {\bibfnamefont {G.}~\bibnamefont
  {Rainovski}}, \bibinfo {author} {\bibfnamefont {P.}~\bibnamefont {Petkov}},
  \bibinfo {author} {\bibfnamefont {J.}~\bibnamefont {Jolie}}, \bibinfo
  {author} {\bibfnamefont {N.}~\bibnamefont {Pietralla}}, \bibinfo {author}
  {\bibfnamefont {A.}~\bibnamefont {Blazhev}}, \bibinfo {author} {\bibfnamefont
  {A.}~\bibnamefont {Damyanova}}, \bibinfo {author} {\bibfnamefont
  {M.}~\bibnamefont {Danchev}}, \bibinfo {author} {\bibfnamefont
  {A.}~\bibnamefont {Dewald}}, \bibinfo {author} {\bibfnamefont
  {C.}~\bibnamefont {Fransen}}, \bibinfo {author} {\bibfnamefont
  {M.}~\bibnamefont {Hackstein}}, \bibinfo {author} {\bibfnamefont
  {D.}~\bibnamefont {Karagyozov}}, \bibinfo {author} {\bibfnamefont
  {O.}~\bibnamefont {M\"oller}}, \bibinfo {author} {\bibfnamefont
  {T.}~\bibnamefont {Pissulla}}, \bibinfo {author} {\bibfnamefont
  {M.}~\bibnamefont {Reese}}, \bibinfo {author} {\bibfnamefont
  {W.}~\bibnamefont {Rother}},\ and\ \bibinfo {author} {\bibfnamefont
  {R.}~\bibnamefont {Topchiyska}},\ }\href
  {https://doi.org/10.1103/PhysRevC.82.037302} {\bibfield  {journal} {\bibinfo
  {journal} {Phys. Rev. C}\ }\textbf {\bibinfo {volume} {82}},\ \bibinfo
  {pages} {037302} (\bibinfo {year} {2010})}\BibitemShut {NoStop}%
\bibitem [{\citenamefont {Pietralla}\ \emph {et~al.}(1999)\citenamefont
  {Pietralla}, \citenamefont {Fransen}, \citenamefont {Belic}, \citenamefont
  {von Brentano}, \citenamefont {Frie\ss{}ner}, \citenamefont {Kneissl},
  \citenamefont {Linnemann}, \citenamefont {Nord}, \citenamefont {Pitz},
  \citenamefont {Otsuka}, \citenamefont {Schneider}, \citenamefont {Werner},\
  and\ \citenamefont {Wiedenh\"over}}]{mo}%
  \BibitemOpen
  \bibfield  {author} {\bibinfo {author} {\bibfnamefont {N.}~\bibnamefont
  {Pietralla}}, \bibinfo {author} {\bibfnamefont {C.}~\bibnamefont {Fransen}},
  \bibinfo {author} {\bibfnamefont {D.}~\bibnamefont {Belic}}, \bibinfo
  {author} {\bibfnamefont {P.}~\bibnamefont {von Brentano}}, \bibinfo {author}
  {\bibfnamefont {C.}~\bibnamefont {Frie\ss{}ner}}, \bibinfo {author}
  {\bibfnamefont {U.}~\bibnamefont {Kneissl}}, \bibinfo {author} {\bibfnamefont
  {A.}~\bibnamefont {Linnemann}}, \bibinfo {author} {\bibfnamefont
  {A.}~\bibnamefont {Nord}}, \bibinfo {author} {\bibfnamefont {H.~H.}\
  \bibnamefont {Pitz}}, \bibinfo {author} {\bibfnamefont {T.}~\bibnamefont
  {Otsuka}}, \bibinfo {author} {\bibfnamefont {I.}~\bibnamefont {Schneider}},
  \bibinfo {author} {\bibfnamefont {V.}~\bibnamefont {Werner}},\ and\ \bibinfo
  {author} {\bibfnamefont {I.}~\bibnamefont {Wiedenh\"over}},\ }\href
  {https://doi.org/10.1103/PhysRevLett.83.1303} {\bibfield  {journal} {\bibinfo
   {journal} {Phys. Rev. Lett.}\ }\textbf {\bibinfo {volume} {83}},\ \bibinfo
  {pages} {1303} (\bibinfo {year} {1999})}\BibitemShut {NoStop}%
\bibitem [{\citenamefont {Pietralla}\ \emph {et~al.}(2001)\citenamefont
  {Pietralla}, \citenamefont {Barton}, \citenamefont {Kr\"ucken}, \citenamefont
  {Beausang}, \citenamefont {Caprio}, \citenamefont {Casten}, \citenamefont
  {Cooper}, \citenamefont {Hecht}, \citenamefont {Newman}, \citenamefont
  {Novak},\ and\ \citenamefont {Zamfir}}]{ru}%
  \BibitemOpen
  \bibfield  {author} {\bibinfo {author} {\bibfnamefont {N.}~\bibnamefont
  {Pietralla}}, \bibinfo {author} {\bibfnamefont {C.~J.}\ \bibnamefont
  {Barton}}, \bibinfo {author} {\bibfnamefont {R.}~\bibnamefont {Kr\"ucken}},
  \bibinfo {author} {\bibfnamefont {C.~W.}\ \bibnamefont {Beausang}}, \bibinfo
  {author} {\bibfnamefont {M.~A.}\ \bibnamefont {Caprio}}, \bibinfo {author}
  {\bibfnamefont {R.~F.}\ \bibnamefont {Casten}}, \bibinfo {author}
  {\bibfnamefont {J.~R.}\ \bibnamefont {Cooper}}, \bibinfo {author}
  {\bibfnamefont {A.~A.}\ \bibnamefont {Hecht}}, \bibinfo {author}
  {\bibfnamefont {H.}~\bibnamefont {Newman}}, \bibinfo {author} {\bibfnamefont
  {J.~R.}\ \bibnamefont {Novak}},\ and\ \bibinfo {author} {\bibfnamefont
  {N.~V.}\ \bibnamefont {Zamfir}},\ }\href
  {https://doi.org/10.1103/PhysRevC.64.031301} {\bibfield  {journal} {\bibinfo
  {journal} {Phys. Rev. C}\ }\textbf {\bibinfo {volume} {64}},\ \bibinfo
  {pages} {031301} (\bibinfo {year} {2001})}\BibitemShut {NoStop}%
\bibitem [{\citenamefont {Ahn}\ \emph {et~al.}(2009)\citenamefont {Ahn},
  \citenamefont {Coquard}, \citenamefont {Pietralla}, \citenamefont
  {Rainovski}, \citenamefont {Costin}, \citenamefont {Janssens}, \citenamefont
  {Lister}, \citenamefont {Carpenter}, \citenamefont {Zhu},\ and\ \citenamefont
  {Heyde}}]{ahn}%
  \BibitemOpen
  \bibfield  {author} {\bibinfo {author} {\bibfnamefont {T.}~\bibnamefont
  {Ahn}}, \bibinfo {author} {\bibfnamefont {L.}~\bibnamefont {Coquard}},
  \bibinfo {author} {\bibfnamefont {N.}~\bibnamefont {Pietralla}}, \bibinfo
  {author} {\bibfnamefont {G.}~\bibnamefont {Rainovski}}, \bibinfo {author}
  {\bibfnamefont {A.}~\bibnamefont {Costin}}, \bibinfo {author} {\bibfnamefont
  {R.}~\bibnamefont {Janssens}}, \bibinfo {author} {\bibfnamefont
  {C.}~\bibnamefont {Lister}}, \bibinfo {author} {\bibfnamefont
  {M.}~\bibnamefont {Carpenter}}, \bibinfo {author} {\bibfnamefont
  {S.}~\bibnamefont {Zhu}},\ and\ \bibinfo {author} {\bibfnamefont
  {K.}~\bibnamefont {Heyde}},\ }\href
  {https://doi.org/https://doi.org/10.1016/j.physletb.2009.06.066} {\bibfield
  {journal} {\bibinfo  {journal} {Phys. Lett. B}\ }\textbf {\bibinfo {volume}
  {679}},\ \bibinfo {pages} {19} (\bibinfo {year} {2009})}\BibitemShut
  {NoStop}%
\bibitem [{\citenamefont {Vanhoy}\ \emph {et~al.}(1995)\citenamefont {Vanhoy},
  \citenamefont {Anthony}, \citenamefont {Haas}, \citenamefont {Benedict},
  \citenamefont {Meehan}, \citenamefont {Hicks}, \citenamefont {Davoren},\ and\
  \citenamefont {Lundstedt}}]{vanhoy1995}%
  \BibitemOpen
  \bibfield  {author} {\bibinfo {author} {\bibfnamefont {J.~R.}\ \bibnamefont
  {Vanhoy}}, \bibinfo {author} {\bibfnamefont {J.~M.}\ \bibnamefont {Anthony}},
  \bibinfo {author} {\bibfnamefont {B.~M.}\ \bibnamefont {Haas}}, \bibinfo
  {author} {\bibfnamefont {B.~H.}\ \bibnamefont {Benedict}}, \bibinfo {author}
  {\bibfnamefont {B.~T.}\ \bibnamefont {Meehan}}, \bibinfo {author}
  {\bibfnamefont {S.~F.}\ \bibnamefont {Hicks}}, \bibinfo {author}
  {\bibfnamefont {C.~M.}\ \bibnamefont {Davoren}},\ and\ \bibinfo {author}
  {\bibfnamefont {C.~L.}\ \bibnamefont {Lundstedt}},\ }\href
  {https://doi.org/doi.org/10.1103/PhysRevC.52.2387} {\bibfield  {journal}
  {\bibinfo  {journal} {Phys. Rev. C}\ }\textbf {\bibinfo {volume} {52}},\
  \bibinfo {pages} {2387} (\bibinfo {year} {1995})}\BibitemShut {NoStop}%
\bibitem [{\citenamefont {Hicks}\ \emph {et~al.}(1998)\citenamefont {Hicks},
  \citenamefont {Davoren}, \citenamefont {Faulkner},\ and\ \citenamefont
  {Vanhoy}}]{hicks1998}%
  \BibitemOpen
  \bibfield  {author} {\bibinfo {author} {\bibfnamefont {S.~F.}\ \bibnamefont
  {Hicks}}, \bibinfo {author} {\bibfnamefont {C.~M.}\ \bibnamefont {Davoren}},
  \bibinfo {author} {\bibfnamefont {W.~M.}\ \bibnamefont {Faulkner}},\ and\
  \bibinfo {author} {\bibfnamefont {J.~R.}\ \bibnamefont {Vanhoy}},\ }\href
  {https://doi.org/doi.org/10.1103/PhysRevC.57.2264} {\bibfield  {journal}
  {\bibinfo  {journal} {Phys. Rev. C}\ }\textbf {\bibinfo {volume} {57}},\
  \bibinfo {pages} {2264} (\bibinfo {year} {1998})}\BibitemShut {NoStop}%
\bibitem [{\citenamefont {Cowan}\ \emph {et~al.}(2021)\citenamefont {Cowan},
  \citenamefont {Sneden}, \citenamefont {Lawler}, \citenamefont {Aprahamian},
  \citenamefont {Wiescher}, \citenamefont {Langanke}, \citenamefont
  {Mart\'{\i}nez-Pinedo},\ and\ \citenamefont
  {Thielemann}}]{RevModPhys.93.015002}%
  \BibitemOpen
  \bibfield  {author} {\bibinfo {author} {\bibfnamefont {J.~J.}\ \bibnamefont
  {Cowan}}, \bibinfo {author} {\bibfnamefont {C.}~\bibnamefont {Sneden}},
  \bibinfo {author} {\bibfnamefont {J.~E.}\ \bibnamefont {Lawler}}, \bibinfo
  {author} {\bibfnamefont {A.}~\bibnamefont {Aprahamian}}, \bibinfo {author}
  {\bibfnamefont {M.}~\bibnamefont {Wiescher}}, \bibinfo {author}
  {\bibfnamefont {K.}~\bibnamefont {Langanke}}, \bibinfo {author}
  {\bibfnamefont {G.}~\bibnamefont {Mart\'{\i}nez-Pinedo}},\ and\ \bibinfo
  {author} {\bibfnamefont {F.-K.}\ \bibnamefont {Thielemann}},\ }\href
  {https://doi.org/10.1103/RevModPhys.93.015002} {\bibfield  {journal}
  {\bibinfo  {journal} {Rev. Mod. Phys.}\ }\textbf {\bibinfo {volume} {93}},\
  \bibinfo {pages} {015002} (\bibinfo {year} {2021})}\BibitemShut {NoStop}%
\end{thebibliography}

%

\end{document}